\documentclass[12pt]{article}

\usepackage[DIV13]{typearea}
\usepackage{pdflscape}
\usepackage{amsfonts, amsmath, amssymb, textcomp}
\usepackage[a4paper,top=3.5cm,bottom=2.8cm,left=2.3cm,right=2.3cm,bindingoffset=5mm]{geometry}
\usepackage{graphicx}
\usepackage{cite}
\usepackage{bbold}
\usepackage{booktabs}

\newcommand{\be}{\beta}

\def\be{\begin{equation}}
\def\ee{\end{equation}}
\def\beq{\begin{equation}}
\def\eeq{\end{equation}}
\def\bc{\begin{center}}
\def\ec{\end{center}}
\def\bea{\begin{eqnarray}}
\def\eea{\end{eqnarray}}

\begin{document}
\begin{titlepage}
\vspace*{-1cm}
\phantom{hep-ph/***}
\flushright
\hfil{DFPD-2013/TH/11}
\hfil{SISSA  28/2013/FISI}\\

\vskip 1.5cm
\begin{center}
\mathversion{bold}
{\LARGE\bf Mixing Patterns from  the Groups $\Sigma (n \varphi)$}\\[3mm]
\mathversion{normal}
\vskip .3cm
\end{center}
\vskip 0.5  cm
\begin{center}
{\large Claudia Hagedorn}~$^{a),b),c)}$,
{\large Aurora Meroni}~$^{c),d)}$\\[2mm]
{\large and Lorenzo Vitale}~$^{e)}$
\\
\vskip .7cm
{\footnotesize
$^{a)}$~Dipartimento di Fisica e Astronomia `G.~Galilei', Universit\`a di Padova,\\ Via Marzolo~8, I-35131 Padua, Italy
\vskip .1cm
$^{b)}$~INFN, Sezione di Padova, Via Marzolo~8, I-35131 Padua, Italy
\vskip .1cm
$^{c)}$~SISSA, Via Bonomea 265, I-34136 Trieste, Italy
\vskip .1cm
$^{d)}$~INFN, Sezione di Trieste, I-34126 Trieste, Italy
\vskip .1cm
$^{e)}$~Institut de Th\'eorie des Ph\'enom\`enes Physiques, EPFL, \\ Lausanne, Switzerland
\vskip .5cm
\begin{minipage}[l]{.9\textwidth}
\begin{center} 
\textit{E-mail:} 
\tt{hagedorn@pd.infn.it}, \tt{ameroni@sissa.it}, \tt{lorenzo.vitale@epfl.ch}
\end{center}
\end{minipage}
}
\end{center}
\vskip 1cm
\begin{abstract}
We survey the mixing patterns which can be derived from the discrete groups $\Sigma (36 \times 3)$, $\Sigma (72 \times 3)$,
$\Sigma (216 \times 3)$ and $\Sigma (360 \times 3)$, if these are broken to abelian subgroups $G_e$ and $G_\nu$ in the charged
lepton and neutrino sector, respectively. Since only $\Sigma (360 \times 3)$ possesses Klein subgroups, only this group allows 
neutrinos to be Majorana particles.
We find a few patterns that can agree well with the experimental data on lepton mixing  in scenarios with small corrections and that predict the reactor mixing angle $\theta_{13}$
to be $0.1 \lesssim \theta_{13} \lesssim 0.2$. All these patterns lead to a trivial Dirac phase. 
Patterns which instead reveal CP violation tend to accommodate the data not well.
We also comment on the outer automorphisms of the discussed groups, since they can be useful for relating inequivalent representations of these groups.
\end{abstract}
\end{titlepage}
\setcounter{footnote}{0}

\section{Introduction}
\label{intro}

Finite discrete non-abelian flavor symmetries have been widely used in order to describe the three lepton mixing angles, for reviews see \cite{reviews,review_math}, whose best fit values and 
$1 \, \sigma$ errors are
\be
\label{eq:global}
\!\sin^2 \theta_{12}= 0.302^{+0.013}_{-0.012} \; , \; \sin^2 \theta_{23}= 0.413 ^{+0.037}_{-0.025} \;\; \mbox{and} \;\; 0.594 ^{+0.021}_{-0.022}  \; , \;
\sin^2 \theta_{13}= 0.0227^{+0.0023}_{-0.0024}
\ee
according to \cite{global_latest}.\footnote{Throughout this paper we use the values given in the left table of table 1 of \cite{global_latest} which are obtained
by performing a global fit in which the reactor fluxes are left free and short baseline reactor data with $L \lesssim 100$ m are included, called ``Free Fluxes + RSBL'' in \cite{global_latest}. For other global fits
which reach similar results see \cite{global_other}.}
A particularly interesting approach is based on the assumption that a (lepton) flavor symmetry $G_l$ is broken to two abelian subgroups $G_e$ and $G_\nu$ in the
charged lepton and neutrino sector, respectively \cite{Gfnontrivial,modular}. The mismatch of the embedding of these two subgroups into the group $G_l$ determines the lepton mixing encoded in the Pontecorvo-Maki-Nakagawa-Sakata
(PMNS) mixing matrix. More precisely, the PMNS matrix is determined up to possible permutations of rows and columns (and phases) so that the results for the mixing angles belong to a small set of
solutions. At the same time, the Jarlskog invariant \cite{jcp} is determined up to a sign (and consequently the Dirac phase $\delta$ can be predicted).
 Left-handed leptons are assigned to an irreducible (faithful) three-dimensional representation of the group $G_l$ in order to achieve at least two non-vanishing mixing angles. As regards the symmetries, $G_e$
 can be any abelian group which is capable of distinguishing among the three generations, while the nature of neutrinos decides on $G_\nu$. For Dirac neutrinos $G_\nu$ is subject to the same constraints as $G_e$, while 
 Majorana neutrinos require $G_\nu$ to be (a subgroup of) a Klein group. A prime example of this approach is tri-bimaximal (TB) mixing \cite{TB}, $\sin^2 \theta_{12}=1/3$, $\sin^2 \theta_{23}=1/2$ and $\theta_{13}=0$, which can
 be derived with the help of  the permutation symmetry $S_4$ \cite{S4TB}. The vanishing of the reactor mixing angle $\theta_{13}=0$, however, is not (well-)compatible with experimental data, since
measurements have shown $\theta_{13} \approx 0.15$ \cite{theta13_exp}. This fact has triggered an intense search for discrete groups which can give rise to mixing patterns with non-vanishing $\theta_{13}$, see e.g. \cite{D96_D384,modular,groupscan1,groupscan2,D6n2_others,groupscan3}, 
as well as it has led to modifications of this approach, e.g. scenarios in which the symmetry of the neutrino sector is only partly contained in the group $G_l$ \cite{GnuZ2_acc} or setups in which the flavor symmetry is combined with a 
CP symmetry and the neutrino sector is invariant under a subgroup of $G_l$ and under the CP symmetry \cite{flavorCP_1,flavorCP_2}.\footnote{For further literature and models with a flavor and a CP symmetry see \cite{flavorCPfurther}.}
 
 We follow the approach to search for new discrete groups and 
  focus in the present paper on the so-called ``exceptional" finite groups $\Sigma (n \varphi)$: $\Sigma (36 \varphi)$, $\Sigma (72 \varphi)$, $\Sigma (216 \varphi)$, $\Sigma (360 \varphi)$ with $\varphi=1, 3$,
 which are subgroups of $SU(3)$ ($\varphi=3$) or of $SU(3)/C$ ($\varphi=1$) with $C$ being the center of $SU(3)$ \cite{Sigmanphi}. 
  Groups with $\varphi=1$ are not suitable for our purposes,\footnote{The group $\Sigma (36)$ is among the finite groups which can
 appear as symmetry group of the scalar sector of a three Higgs doublet model without leading to a potential with a continuous symmetry \cite{3HDMdisc}.} since they do not possess 
 an irreducible faithful three-dimensional representation, while the groups with $\varphi=3$ do have such representations. We compute all mixing patterns
related to these groups, assuming neutrinos to be either Dirac or Majorana particles.\footnote{This is in contrast to the recently performed scans of groups with an order smaller than 200 \cite{groupscan3},
an order smaller than 512 \cite{groupscan1} or an order smaller than 1536 \cite{groupscan2} that had the scope to only figure out the groups
which, if broken in a non-trivial way, lead to mixing patterns, accommodating best the experimental data \cite{global_latest}.} 
 Among these four groups only $\Sigma (360 \times 3)$ has Klein subgroups and is thus suitable as $G_l$, if neutrinos are assumed to be Majorana particles.
$\Sigma (36 \times 3)$, $\Sigma (72 \times 3)$ and $\Sigma (216 \times 3)$ instead are only appropriate 
as $G_l$, if neutrinos are Dirac particles or only one of the $Z_2$ symmetries of the neutrino mass matrix is contained in $G_l$. In the latter case only
one of the columns of the PMNS matrix is determined through the breaking of the flavor symmetry $G_l$ to $G_e$ and to $G_\nu$. One can generalize this idea and consider as $G_e$ or $G_\nu$
 subgroups of $G_l$ which are generated by elements that are represented by matrices with two degenerate eigenvalues. Then only one of the rows or columns of the PMNS matrix is fixed.
 This row/column is called ``mixing vector" \cite{groupscan1}. 
 
The authors of \cite{dMFV} have shown that the groups $\Sigma (72 \times 3)$, $\Sigma (216 \times 3)$ and $\Sigma (360 \times 3)$
 are suitable for realizing a version of minimal flavor violation in the quark sector
with a discrete group, since the products of (faithful) three-dimensional representations 
with their complex conjugates decompose in the same way as in $SU(3)$, i.e. ${\bf 3} \times \bf{3^\star} = {\bf 1}+ {\bf 8}$ with ${\bf 8}$ being irreducible in $G_l$.
The group $\Sigma (168) \simeq PSL (2,Z_7)$ has also been recognized as interesting by the authors of \cite{dMFV}. However, the possible mixing patterns arising from this group have been already analyzed in \cite{modular}, mainly under
the assumption that neutrinos are Majorana particles.\footnote{For the sixth group of this type, called $\Sigma (60)$ and isomorphic to $A_5$, the mixing patterns have also already been studied in detail for Majorana neutrinos with
some comments on Dirac neutrinos \cite{A4S4A5patterns,modular}. This group however is not so interesting according to \cite{dMFV}, since it permits several more invariants than the groups associated with discrete minimal flavor
violation in the quark sector.}

We find that the two smaller groups $\Sigma (36 \times 3)$ and $\Sigma (72 \times 3)$ only give rise to a few new mixing patterns, while the larger groups lead to several new patterns, since they have a richer structure of subgroups.
Generically one sees (as already suggested by the scans of groups making use of the computer program GAP \cite{groupscan1,groupscan2,groupscan3}) that only very few patterns are well-compatible
with the experimental data of all three different mixing angles \cite{global_latest}, if only small corrections are admitted, i.e. for each group one to two. Such small corrections, typically of order $\lambda^2 \approx 0.04$, are usually expected 
to arise in concrete models, see e.g. \cite{reviews}, and thus any pattern which can be brought into agreement with the experimental results including this type of corrections is considered as very interesting for model building.
However, if one of the other patterns is realized in an explicit model one either has to invoke large corrections to the leading
order results in order to reconcile them with the data or one needs to consider modifications of the breaking pattern, such as involving a CP symmetry \cite{flavorCP_1,flavorCP_2}. It is noteworthy that patterns which fit 
the experimental data better usually do not lead to a non-trivial Dirac phase (although all mixing angles are different from 0 and $\pi/2$).\footnote{Similar observations can also be made in the analyses found in \cite{groupscan2,D6n2_others,groupscan3}.}
 On the other hand, among the patterns associated with the group $\Sigma (216 \times 3)$ we find several ones which give rise to non-vanishing CP violation and at the same time accommodate the solar and 
 atmospheric mixing angles at the $3  \, \sigma$ level or better. Only the reactor mixing angle turns out to be too large, see table \ref{tab:Sigma216x3Pattern2}. Furthermore, we apply outer automorphisms of the groups
 $\Sigma (n \varphi)$, $\varphi=3$, in order to relate their irreducible faithful three-dimensional representations among each other. In this way, we see that it is sufficient to focus on a single 
 three-dimensional representation in the discussion of the mixing patterns.
 
The paper is structured as follows: in Section \ref{sec2} we review the approach used in order to relate the mixing pattern with the group $G_l$ and its breaking to the subgroups $G_e$ and $G_\nu$ in the charged
lepton and neutrino sectors. In Section \ref{sec3} we discuss the mathematical properties of the different groups $\Sigma (n \varphi)$, the representations to which the three generations of left-handed leptons can be 
assigned, the relevant abelian subgroups and the resulting mixing patterns and predictions for mixing angles $\theta_{ij}$ and the Jarlskog invariant $J_{CP}$. We use as figure of merit $\chi^2$ assuming the
best fit values and $1 \, \sigma$ errors of the mixing angles as given in eq.(\ref{eq:global}). We also comment on the mixing vectors for each group.
 We summarize in Section \ref{concl}. Comments on outer automorphisms of the groups $\Sigma (36 \times 3)$, $\Sigma (72 \times 3)$ and $\Sigma (216 \times 3)$ can be found in Appendix \ref{app}.

\section{Approach}
\label{sec2}

We briefly review how lepton mixing can be predicted with a non-abelian discrete flavor symmetry $G_l$ which is broken, spontaneously or explicitly, to two different
abelian subgroups $G_e$ and $G_\nu$ in the charged lepton and neutrino sector \cite{Gfnontrivial,modular}. We assign the three generations of left-handed leptons to an irreducible three-dimensional 
representation, because we want to discuss patterns with at least two non-vanishing lepton mixing angles. We furthermore choose the representation to be faithful in the group $G_l$ 
so that we (mostly) study mixing patterns which originate from this group itself and not from one of its proper
subgroups.\footnote{The requirement of faithfulness of the three-dimensional representation also excludes the possibility to choose $G_e$ or $G_\nu$ to be non-abelian,
since a faithful representation of $G_l$ usually decomposes into an irreducible representation of dimension larger than one in non-abelian $G_e$ or $G_\nu$
so that a distinction among the three lepton generations becomes impossible.} In order to fix the mixing angles 
we also have to assume that the three generations of left-handed leptons transform as three inequivalent one-dimensional
representations under the residual symmetries $G_e$ and $G_\nu$. A possible set of generators of $G_e$ and $G_\nu$ is 
called $g_{e,i}$, $i=1,2, ...$ and $g_{\nu,j}$, $j=1,2,...$, respectively.\footnote{In abuse of notation we denote in the following the abstract
elements of the groups $G_e$, $G_\nu$ and $G_l$ with the same symbol as their matrix representatives.} Clearly, these are also elements of the group $G_l$. 
The requirement that a subgroup $G_e$ is conserved   
 entails that the combination $m_e m_e^\dagger$ (with the convention $\overline{\Psi}_L m_e \Psi_R$) has to be invariant under the symmetry $G_e$, i.e.
 \be
\label{eq:memedag}
g_{e,i}^\dagger m_e m_e^\dagger g_{e,i} = m_e m_e^\dagger \;\;\; , \;\; i=1,2,... \, ,
\ee
  while requiring that $G_\nu$ is a symmetry of the neutrino mass matrix $m_\nu$ entails that $m_\nu m_\nu^\dagger$ ($m_\nu$ itself) is invariant under $G_\nu$ for Dirac (Majorana) neutrinos 
 \be
\label{eq:mnu}
g_{\nu,j}^\dagger m_\nu m_\nu^\dagger g_{\nu,j} = m_\nu m_\nu^\dagger \;\;\; , \;\; j=1,2,... \;\;\; \mbox{or} \;\;\; g_{\nu,j}^T m_\nu g_{\nu,j} = m_\nu \;\;\; , \;\; j=1,2,... \, .
\ee
 For Majorana neutrinos $G_\nu$ is strongly constrained, since it has to be (a subgroup of) a Klein symmetry for three generations
 of neutrinos. If neutrinos are Dirac particles, $G_\nu$ is subject to the same constraint as $G_e$ of the charged lepton sector, i.e. it can be any abelian group, capable of distinguishing the three generations. In our actual analysis we follow
a minimality principle when choosing $G_e$ and $G_\nu$, since we consider all minimal (abelian) subgroups of the group $G_l$ which allow the distinction of three generations.\footnote{For example, if a 
subgroup $G_e=G_1$ is sufficient in order to distinguish among the generations, we do not discuss the case in which $G_e=G_1 \times G_2 \subset G_l$,
since the presence of the group $G_2 \subset G_e$ would not add any information as far as lepton mixing is concerned.} Furthermore, we wish to focus on mixing patterns
which are associated with the group $G_l$ and not only with one of its subgroups and thus we require that the generators of the subgroups $G_e$ and $G_\nu$
give rise to the entire group $G_l$. 
We can change basis via the unitary transformations $U_e$ and $U_\nu$ so that $g_{e,i}$ and $g_{\nu,j}$ are diagonalized ($g_{e\, (\nu), i \, (j)}^d$ are diagonal matrices), respectively, 
\be
\label{eq:UeUnu}
U_e^\dagger g_{e,i} U_e = g_{e,i}^d \;\;\; , \;\; i=1,2,... \;\;\; \mbox{and} \;\;\; U_\nu^\dagger g_{\nu,j} U_\nu = g_{\nu,j}^d \;\;\; , \;\; j=1,2,... \, .
\ee
Then, also $m_e m_e^\dagger$ and $m_\nu m_\nu^\dagger$ for Dirac (or $m_\nu$ for Majorana) neutrinos are diagonalized by $U_e$ and $U_\nu$, respectively, and
the PMNS mixing matrix is derived as
\be
\label{eq:PMNS}
U_{PMNS} = U_e^\dagger U_\nu \; .
\ee
The matrices $U_e$ and $U_\nu$ are uniquely determined up to permutations and phases of their column vectors, since we consider a scenario in which the three generations of left-handed
leptons are distinguished, but no further assumption or prediction is made about the lepton masses. 
As a consequence, $U_{PMNS}$ is fixed up to permutations of rows and columns and the multiplication with
phase matrices from the left- and the right-hand side. In turn, the mixing angles $\theta_{ij}$ which only depend on the absolute values of the elements of the PMNS matrix $||U_{PMNS}||$, and 
the Jarlskog invariant $J_{CP}$ which can be expressed as $J_{CP}= \mathrm{Im} (U_{PMNS,11} U_{PMNS,13}^\star U_{PMNS,31}^\star U_{PMNS,33})$,
take only values which belong to a small set (see \cite{pdg} for our convention of mixing angles and CP phases).
 Note that an exchange of the roles of the subgroups $G_e$ and $G_\nu$ transforms the PMNS matrix into its hermitian conjugate.
Two pairs of subgroups, $\{ G_e, G_\nu \}$ and $\{ G'_e, G'_\nu \}$, lead to the same result for the PMNS matrix, if their generators $g_{e,i}$, $g_{\nu,j}$ and $g'_{e,i}$, $g'_{\nu,j}$ are related
by a similarity transformation, i.e. if these pairs of groups are conjugate. Notice that two pairs $\{ G_e, G_\nu \}$ and $\{ G'_e, G'_\nu \}$ which are only isomorphic, but not conjugate, may or may not lead to 
different mixing patterns. In the case of the group $\Sigma (36 \times 3)$, for example, a combination $\{ G_e, G_\nu \}$ of a certain type of subgroups gives, if admissible, always
rise to the same mixing pattern, even if not all pairs  $\{ G_e, G_\nu \}$ are related by similarity transformations. In the case  $G_l= \Sigma (216 \times 3)$, $G_e=Z_4$ and $G_\nu=Z_{18}$, see
table \ref{tab:Sigma216x3Pattern1} and \ref{tab:Sigma216x3Pattern2}, instead it is not sufficient to only specify the type of subgroups, since  
the combination $G_e=Z_4$ and $G_\nu=Z_{18}$ can give rise to pattern IV as well as to patterns IXa and IXb. If we exchange the roles of $G_e$ and $G_\nu$, i.e. $G_e=Z_{18}$ and $G_\nu=Z_4$, we see
that also the patterns Ia and Ib in table \ref{tab:Sigma216x3Pattern1} originate from the combination of these subgroups.

We also consider the case in which the generators of $G_e$ or $G_\nu$ are represented by matrices with two degenerate eigenvalues. In this case only one generation can be distinguished from the other two
ones and thus only one row or column can be fixed through the choice of $G_e$, $G_\nu$ and $G_l$. Such a row or column is called ``mixing vector" \cite{groupscan1}. 

Up to this point we have not fully specified the three-dimensional representation as which the three generations of left-handed leptons transform, apart from requesting that it should be irreducible and faithful in $G_l$.
However, in general a group $G_l$ possesses several representations which meet these requirements. If so, we have to analyze whether the results for the mixing patterns do also depend on the choice of
the three-dimensional representation and not only on the choice of the groups $G_e$, $G_\nu$ and $G_l$.
We will show that in all groups $\Sigma (n \varphi)$ the mixing patterns are independent of the actual choice of the representation and that it is thus sufficient to study
the mixing patterns for only one of them. In doing this, outer automorphisms are particularly useful, since they allow to relate inequivalent representations of a group.

\mathversion{bold}
\section{Groups $\Sigma (n \varphi)$ and mixing patterns}
\mathversion{normal}
\label{sec3}

The four groups we discuss in the following are $\Sigma (36 \times 3)$, $\Sigma (72 \times 3)$, $\Sigma (216 \times 3)$ and $\Sigma (360 \times 3)$.
As we show, only the latter group has Klein subgroups, while the three other ones not. Thus, if $G_\nu$ is required to be a subgroup of $G_l$
and neutrinos are Majorana particles, only $\Sigma (360 \times 3)$ can be chosen as $G_l$. In the other cases neutrinos are either Dirac particles
or $G_\nu$ is only partly contained in $G_l$ (i.e. one of the two $Z_2$ factors of the Klein group is accidental).

We first present a set of generators for each group together with the relations they have to satisfy. We then focus on the irreducible three-dimensional
representations and discuss in some detail why it is sufficient for each of the four groups to only consider one of the representations when deriving the 
lepton mixing patterns. We detail the different abelian subgroups of the different $\Sigma (n \varphi)$ with $\varphi=3$ and comment on those
which are admissible as $G_e$ and $G_\nu$, because they are capable of distinguishing the three lepton generations (in our chosen three-dimensional
representation), and on those which can only be employed in order to predict a mixing vector. 
As we will see, the groups $\Sigma(36 \times 3)$,
$\Sigma (72 \times 3)$ and $\Sigma (360 \times 3)$ contain $Z_2$ and $Z_6$ generating elements which are represented by matrices with partly degenerate eigenvalues, while $\Sigma (216 \times 3)$
also comprises $Z_9$ generating elements with this property.

We show all patterns which can
arise from a certain choice of $G_l$, $G_e$ and $G_\nu$, following our minimality principle for $G_e$ and $G_\nu$ as well as imposing the requirement
that the generators of these two groups give rise to the whole group $G_l$. As explained, the ordering of rows and columns is not fixed by the choice of the subgroups
$G_e$ and $G_\nu$. We choose it in the presentation of our results in such a way that the value of $\chi^2$ is minimized. The latter is
computed in the standard way and with the experimental results for the mixing angles found in \cite{global_latest} in the left table of table 1. 
 If all three mixing angles are in agreement with the experimental data at the $3 \, \sigma$ level or better, the value of $\chi^2$ is $\chi^2 \lesssim 27$ (the converse is in general not true and for this
reason we always list the values of all three mixing angles).
In several occasions we
do not only show the pattern with rows and columns permuted in a way that $\chi^2$  is minimized, but we also mention further patterns with a
 slightly larger $\chi^2$. The exact notion of ``slightly larger" depends on the minimal value of $\chi^2$: for the latter being smaller than 100 we 
 consider patterns with values of $\chi^2$ up to $5\%$ larger than the minimal one, while for a minimal value of $\chi^2$ between 100 and 1000 
we only mention patterns  with values of $\chi^2$ up to $1\% \div 2 \%$ larger than the minimal one. In all cases we additionally require
that at least one mixing angle agrees within $3 \, \sigma$ with the experimental results \cite{global_latest}. If the minimal value of $\chi^2$ is larger than 1000
we only show the pattern belonging to this value, unless there are two mixing angles which are in accordance with the experimental data within
 $3 \, \sigma$. If there are two permutations of a pattern leading to a similar $\chi^2$ and we display both, we denote them with the same Roman numeral,
 but a different letter, e.g. pattern Ia and Ib, as in table \ref{tab:Sigma216x3Pattern1} for the group $\Sigma (216 \times 3)$. It turns out that in all
 cases the two permutations of a certain pattern are related through the exchange of the second and third rows of the PMNS matrix. Thus, they
 generally lead to the same reactor and solar mixing angles, while the atmospheric one changes from $\theta_{23}$ to $\pi/2 -\theta_{23}$, i.e. 
 $\sin^2 \theta_{23}$ is replaced by $1-\sin^2 \theta_{23}$. 
  The tables \ref{tab:Sigma216x3Pattern1}, \ref{tab:Sigma216x3Pattern2} and \ref{tab:Sigma360x3Pattern1}-\ref{tab:Sigma360x3Pattern3} 
summarize our results for the mixing patterns derived from the groups $\Sigma (216 \times 3)$
and $\Sigma (360 \times 3)$. We mark mixing angles that are in accordance with the experimental data at the $3 \, \sigma$ level or better in these tables with a star ($\star$). 
 For all patterns with $\chi^2 < 100$ we have checked that an agreement of the mixing angles with the experimental data at the $3 \, \sigma$ level
or better can be easily achieved in models with small corrections of order $\lambda^2 \approx 0.04$.\footnote{We explicitly studied a scenario in which, due to an appropriate choice of basis, 
the lepton mixing arises at leading order only from the neutrino sector. 
 Assuming that subleading corrections are present in the charged lepton sector which induce small mixing angles of order $\lambda^2 \approx 0.04$ there, 
all mixing angles can be accommodated within their $3 \, \sigma$ ranges.} 
Thus, we consider all these patterns as phenomenologically interesting.

\mathversion{bold}
\subsection{Group $\Sigma (36 \times 3)$}
\mathversion{normal}
\label{sec31}

The group $\Sigma (36 \times 3)$ (identification number [[108,15]] in the library SmallGroups \cite{smallgroup} which is accessible
via the computer algebra system GAP \cite{gap}) has 108 elements and can 
be represented in terms of three generators $a$, $v$ and $z$ which fulfill the following relations \cite{Sigma}
\be
\label{eq:genrelS36x3}
a^3=1 \; , \;\; v^4=1\; , \;\; z^3=1 \; , \;\; a v^{-1} z v=1 \; , \;\; a v z^{-1} v^{-1} = 1 \; , \;\; (a z)^3 =1 \; .
\ee
Here and in the following, we denote the neutral element of the group with $1$.
For one of the irreducible faithful three-dimensional representations, identified with ${\bf 3^{(0)}}$ of \cite{Sigma}, the generators
are given by the matrices
\be
\label{eq:genS36x3}
a= \left( \begin{array}{ccc}
 0 & 1 & 0\\
 0 & 0 & 1\\
 1 & 0 & 0
\end{array}
\right) \;\; , \;\;\;
v= \frac{1}{\sqrt{3} i} \, \left( \begin{array}{ccc}
 1 & 1 & 1\\
 1 & \omega & \omega^2\\
 1 & \omega^2 & \omega
\end{array}
\right) \;\; , \;\;\;
z= \left( \begin{array}{ccc}
 1 & 0 & 0\\
 0 & \omega & 0\\
 0 & 0 & \omega^2
\end{array}
\right)
\ee
 with $\omega=e^{2 \pi i/3}$. Note that all generators have determinant +1.
The group $\Sigma (36 \times 3)$ possesses in total eight faithful irreducible three-dimensional representations which form four complex conjugated pairs. 
The representation matrices of the generators $a$, $v$ and $z$ in the three other faithful three-dimensional
representations ${\bf 3^{(p)}}$, $\mathrm p=1,2,3$, are closely related to $a$, $v$ and $z$ of ${\bf 3^{(0)}}$
\be
a \; , \;\; i^{\rm p} v \; , \;\; z \;\;\; \mbox{with} \;\;\; \rm p=1,2,3
\ee
and those of (the complex conjugated repsentations) ${\bf (3^{(p)})^\star}$, $\rm p=0,1,2,3$, are obtained through complex conjugation. 
 Apart from these, the irreducible representations of the group are: four one-dimensional representations, two real and two complex ones, and
two real four-dimensional representations. Notice that they are all unfaithful in $\Sigma (36 \times 3)$. The character table of the group can be found, for example, in \cite{Sigma}.

Clearly, the mixing patterns which can be derived by assuming that the three generations of leptons 
are assigned to any of the representations ${\bf 3^{(p)}}$ have to coincide with those derived from assigning them to ${\bf 3^{(0)}}$, since the additional factor $i^{\rm p}$ in
 the generator $v$ cannot change the eigenvectors. Similarly, all mixing matrices originating from one of the complex conjugated representations  ${\bf (3^{(p)})^\star}$ are the same as those
derived from ${\bf 3^{(p)}}$, up to complex conjugation. In Appendix \ref{appS36x3} we comment on the outer automorphism group of $\Sigma (36 \times 3)$ and the possibility to understand
relations among the three-dimensional representations using the latter.  
The products of two irreducible three-dimensional representations are of the following type
\be
{\bf 3^{(p_1)}} \times {\bf 3^{(p_2)}} =  ({\bf 3^{(p_3)}})^\star +  ({\bf 3^{(p_4)}})^\star +  ({\bf 3^{(p_5)}})^\star \;\;\; \mbox{and} \;\;\;
({\bf 3^{(p_1)}}) \times  ({\bf 3^{(p_2)}})^\star = {\bf 1^{(q)}} + {\bf 4} + {\bf 4'} \; ,
\ee
with $\mathrm{p}_i, \mathrm{q}=0, 1, 2, 3$, $i=1,...,5$, especially
\be
{\bf 3^{(0)}} \times {\bf 3^{(0)}} =  ({\bf 3^{(0)}})^\star +  ({\bf 3^{(1)}})^\star +  ({\bf 3^{(3)}})^\star \;\;\; \mbox{and} \;\;\;
({\bf 3^{(0)}}) \times  ({\bf 3^{(0)}})^\star = {\bf 1^{(0)}} + {\bf 4} + {\bf 4'} \; .
\ee
Thus, this group does not belong to those singled out in the analysis of \cite{dMFV} as being a suitable candidate for the implementation of a version of minimal flavor violation with a discrete group. Nevertheless,
we discuss the mixing patterns which can be derived from $\Sigma (36 \times 3)$, since its order is still rather small.

For analyzing all abelian subgroups of $\Sigma (36 \times 3)$ it is convenient to make use of the fact that all elements $g$ of the group $\Sigma (36 \times 3)$ can be written,  in the three-dimensional representations, in the form
\be
g=\omega^o z^\zeta a^\alpha v^\chi \;\;\; \mbox{with} \;\;\; o, \, \zeta, \, \alpha=0,1,2 \; , \chi=0,1,2,3 
\ee
and that a representative (again in the three-dimensional representations) for each of the fourteen classes $c_1 \, {\cal C}_{c_2}$ (with $c_1$ denoting the number of elements of the class ${\cal C}_{c_2}$ and $c_2$ their order) 
can be found \cite{Sigma} 
\bea
&&1 \, {\cal C}_1 \; : \; 1 \;\; , \;\;\; 9 \, {\cal C}_2 \; : \; v^2 \;\; , \;\;\;  1 \, {\cal C}_3 \; : \; a z a^2 z^2 =\omega \, 1 \;\; , \;\;\;  1 \, {\cal C}_3 \; : \; a^2 z a z^2 =\omega^2 \, 1 \;\; , \;\;\; 
\\ \nonumber
&&12 \, {\cal C}_3 \; : \; z \;\; , \;\;\; 12 \, {\cal C}_3 \; : \; z a \;\; , \;\;\; 9 \, {\cal C}_4 \; : \; v \;\; , \;\;\; 9 \, {\cal C}_4 \; : \; v^3 \;\; , \;\;\; 9 \, {\cal C}_6 \; : \; \omega \, v^2 \;\; , \;\;\; 
\\ \nonumber
&&9 \, {\cal C}_6 \; : \; \omega^2 \, v^2 \;\; , \;\;\;
9 \, {\cal C}_{12} \; : \; \omega \, v \;\; , \;\;\; 9 \, {\cal C}_{12} \; : \; \omega^2 \, v \;\; , \;\;\; 9 \, {\cal C}_{12} \; : \; \omega \, v^3 \;\; , \;\;\; 9 \, {\cal C}_{12} \; : \; \omega^2 \, v^3 \;\; .
\eea
The abelian subgroups of $\Sigma (36 \times 3)$ are nine $Z_2$ symmetries, 13 $Z_3$ symmetries (including $C$, the center of $SU(3)$), nine $Z_4$ and nine $Z_6$ symmetries, as well as
nine $Z_{12}$ groups. Furthermore, there are four groups $Z_3 \times Z_3$ where, however, one of the two $Z_3$ groups is the center of $SU(3)$. We have checked that all $Z_2$ symmetries,
all $Z_4$, all $Z_6$ and also all $Z_{12}$ symmetries are conjugate to each other. The $Z_3$ symmetries fall instead into three categories: two with six such symmetries which are conjugate among each other
and a third one which comprises only the center of $SU(3)$. Since $\Sigma (36 \times 3)$ does not contain a Klein group,
neutrinos have to be either Dirac particles or the symmetry of the neutrino sector is only partly contained in $G_l$. Using the representation matrices in eq.(\ref{eq:genS36x3}) we see that
30 elements of $\Sigma (36 \times 3)$ are represented by matrices with (at least) two degenerate eigenvalues: the ones belonging to the center of $SU(3)$, the ones generating $Z_2$ groups as well as
the ones generating $Z_6$ symmetries. Therefore these elements cannot be used
as generators of the subgroups $G_e$ and $G_\nu$ alone, if we wish to fix all three mixing angles through the breaking of $G_l$ to $G_e$ and to $G_\nu$. 

With this classification of subgroups of $\Sigma (36 \times 3)$ at hand we can distinguish the following cases  in which the breaking of the group $\Sigma (36 \times 3)$ to $G_e$ and to $G_\nu$ fixes the mixing pattern: 
$G_e=Z_3$ and $G_\nu=Z_3$, $G_e=Z_3$ and $G_\nu=Z_4$ or $G_\nu=Z_{12}$,
 $G_e=Z_4$ or $G_e=Z_{12}$ and  $G_\nu=Z_4$ or $G_\nu=Z_{12}$. Out of these possibilities only two categories are interesting and can pass our constraint of generating the entire group $\Sigma (36 \times 3)$
with the generators of $G_e$ and $G_\nu$.\footnote{For the choice $G_e=Z_3$ and $G_\nu=Z_3$ we encounter only cases in which the mixing matrix is a permutation of the unit matrix or all its elements have the same
absolute value or one of the subgroups $G_e$ or $G_\nu$ is identified with the center of $SU(3)$ so that the three generations cannot be distinguished.} 
The first interesting case arises for $G_e=Z_3$ and $G_\nu=Z_4$ or $G_\nu=Z_{12}$. For both possibilities, $G_\nu=Z_4$ and $G_\nu=Z_{12}$, there are five 
independent choices of pairs $\{ G_e , G_\nu \}$ whose sets of generators $\{ g_e, g_\nu \}$ are not related among
each other via similarity transformations. For one of them $G_e=Z_3$ is identified with the center of $SU(3)$ so that the three generations cannot be distinguished and thus this choice is not considered. The other
four independent choices of generators instead fulfill all requirements, i.e. their generators allow for a distinction of the three generations and they generate together the original group $\Sigma (36 \times 3)$.
All these admissible choices lead to the mixing pattern
\small
\be
\label{S36x3_1}
||U_{PMNS}||= \frac{1}{\sqrt{2 (3+\sqrt{3})}} \, \left(
\begin{array}{ccc}
1+\sqrt{3} & \sqrt{2} & 0\\
1& \sqrt{2+\sqrt{3}} & \sqrt{3+\sqrt{3}}\\
1& \sqrt{2+\sqrt{3}} & \sqrt{3+\sqrt{3}}
\end{array}
\right) \approx
\left( \begin{array}{ccc}
0.888 & 0.460 & 0\\
0.325 & 0.628 & 0.707\\
0.325 & 0.628 & 0.707
\end{array}
\right) \; .
\ee
\normalsize
The mixing angles read $\sin^2 \theta_{12} \approx 0.211$, $\sin^2 \theta_{23}=0.5$ and $\theta_{13}=0$. Since the reactor mixing angle vanishes, also the Jarlskog invariant $J_{CP}$ vanishes.
The value of $\chi^2$ is $\chi^2 \approx 151.5$ in this case and only the atmospheric mixing angle is in agreement with the experimental results within $3 \, \sigma$.
 Examples of generators of $G_e$ and $G_\nu$ are $g_e= \omega^2 z a^2$ and $g_\nu=z a^2 v$ and $g_\nu =\omega z^2 a^2 v^3$, respectively. The pattern in eq.(\ref{S36x3_1}) has been recently
 found in a study in which the flavor group $\Delta (27)$ is combined with a CP symmetry \cite{Delta27CP}.

If $G_e$ and $G_\nu$ are instead both $Z_4$ or $Z_{12}$ subgroups of $\Sigma (36 \times 3)$, we find that there exist three independent choices of pairs $\{ G_e , G_\nu \}$ which are isomorphic, but not conjugate.
 One of these choices we have to exclude, because $G_e$ and $G_\nu$ are the same or one is a subgroup of the other one, while
the other two choices are admissible. We obtain
for both of them as mixing pattern
\small
\be
\label{S36x3_2}
||U_{PMNS}||= \frac{1}{2 \sqrt{2}} \,  \left(
\begin{array}{ccc}
 \sqrt{3+\sqrt{3}} & \sqrt{3} & \sqrt{2-\sqrt{3}}\\
 \sqrt{2} & \sqrt{3-\sqrt{3}} & \sqrt{3+\sqrt{3}}\\
 \sqrt{3-\sqrt{3}} & \sqrt{2+\sqrt{3}} & \sqrt{3}
\end{array}
\right)\approx
\left(
\begin{array}{ccc}
 0.769 & 0.612 & 0.183 \\
 0.500 & 0.398 & 0.769\\
 0.398 & 0.683 & 0.612
\end{array}
\right)
\ee
\normalsize
leading to the mixing angles $\sin^2 \theta_{12}  \approx 0.388$, $\sin^2 \theta_{23} \approx 0.612$ and $\sin^2 \theta_{13} \approx 0.033$. Again, $J_{CP}$ vanishes.
The value of $\chi^2$ is 69.1. If the second and third rows of the PMNS matrix in eq.(\ref{S36x3_2}) are exchanged, the atmospheric mixing angle reads $\sin^2 \theta_{23}  \approx 0.388$ like the
solar one and the value of $\chi^2$ slightly increases, $\chi^2 \approx 69.4$. In both cases only the atmospheric mixing angle agrees within $3 \, \sigma$ with the experimental data.
Examples for the generators of the groups $G_e$ and $G_\nu$ are: if both are $Z_4$ subgroups, we can take $g_e= z a^2 v$ and $g_\nu= \omega^2 z^2 a v^3$; if $G_e$ is a $Z_4$ and $G_\nu$
a $Z_{12}$ subgroup, we can use $g_e=z a^2 v$ and $g_\nu=\omega z^2 a^2 v^3$ (we might also switch the roles of $G_e$ and $G_\nu$ and still obtain the same mixing pattern); and for $G_e$
and $G_\nu$ being both $Z_{12}$ subgroups, we can choose for example the generators $g_e=\omega z^2 a^2 v^3$ and $g_\nu=\omega^2 z v$. As one can see, all results which can be achieved
with a $Z_4$ symmetry can also be obtained with an appropriate $Z_{12}$ group. This happens, because all $Z_{12}$ generating elements have a matrix representation which can be written as product
of a matrix giving rise to a $Z_4$ generating element and a matrix representing a non-trivial element of the center of $SU(3)$.

Furthermore, we can consider the case in which one of the subgroups $G_e$ or $G_\nu$ is generated by an element of $\Sigma (36 \times 3)$ which has two degenerate eigenvalues, i.e. $G_e$ or $G_\nu$
is a $Z_2$ or $Z_6$ symmetry.  
The resulting mixing vectors
can take only three possible forms (referring to the vectors with entries given in absolute values), namely a vector with only one non-vanishing entry, a vector with two equal entries and one vanishing one or a vector equal to the first column 
of the PMNS matrix shown in eq.(\ref{S36x3_2}). It is not by chance that the set of mixing vectors arising from $Z_2$ and $Z_6$ generating elements is the same, since one can check that all $Z_6$ generating
elements are represented by matrices which are products of a matrix representing the generator of a $Z_2$ and a matrix which represents a non-trivial element of the center of $SU(3)$.

\mathversion{bold}
\subsection{Group $\Sigma (72 \times 3)$}
\mathversion{normal}
\label{sec32}

In order to generate the group $\Sigma (72 \times 3)$ which has the identification number [[216,88]] in the library SmallGroups 
we add one generator $x$ to the set of generating elements $a$, $v$ and $z$ of the group $\Sigma (36 \times 3)$. This
additional generator fulfills the relations \cite{Sigma}
 \be
 \label{eq:genrelS72x3}
x^2 = z^2 v^2 \; \;\; \mbox{and}  \;\;\; x^4=1 \; .
\ee
In the three-dimensional representation, called ${\bf 3^{(0,0)}}$ in \cite{Sigma}, the matrices of the
generators can be chosen as $a$, $v$, $z$ in eq.(\ref{eq:genS36x3}) together with
\be
\label{eq:genS72x3}
x= \frac{1}{\sqrt{3} i} \, \left( \begin{array}{ccc}
 1 & 1 & \omega^2\\
 1 & \omega & \omega\\
 \omega & 1 & \omega
\end{array}
\right) \;\; .
\ee
Notice also the matrix representative of the element $x$ has determinant +1.
This representation is one of the eight faithful irreducible three-dimensional representations of $\Sigma (72 \times 3)$ which form four complex conjugated pairs  ${\bf 3^{(p_1,p_2)}}$ and  ${\bf (3^{(p_1,p_2)})^\star}$
with $\rm p_{1,2}=0,1$. The representation matrices of the generators $a$, $v$, $z$ and $x$ in the three-dimensional representations ${\bf 3^{(p_1,p_2)}}$ are given by
\be
a \; , \;\; (-1)^{\mathrm{p_1}} v \; , \;\; (-1)^{\mathrm{p_2}} x \; , \;\; z   \;\;\; \mbox{with} \;\;\; \rm p_{1,2}=0,1
\ee
and those of ${\bf (3^{(p_1,p_2)})^\star}$ by the complex conjugated representation matrices. Like in the case of the three-dimensional representations of the group $\Sigma (36 \times 3)$
this shows that ${\bf 3^{(p_1,p_2)}}$ and $({\bf 3^{(p_1,p_2)}})^\star$, $\rm p_{1,2}=0,1$, all must lead to the same results for the mixing patterns. 
In Appendix \ref{appS72x3} we use the outer automorphisms of $\Sigma (72 \times 3)$ in order
to relate the inequivalent three-dimensional representations. 
Apart from  ${\bf 3^{(p_1,p_2)}}$ and $({\bf 3^{(p_1,p_2)}})^\star$ the group contains four one-dimensional, one two-dimensional, one eight-dimensional
and two six-dimensional irreducible representations. The latter two are complex conjugated and faithful, while the former ones are all real and unfaithful. The character
table of $\Sigma (72 \times 3)$ can be found in \cite{Sigma}.

The generic form of the products among two three-dimensional representations is like in $SU(3)$, i.e.
\be
{\bf 3^{(p_{1,1},p_{2,1})}} \times {\bf 3^{(p_{1,2},p_{2,2})}}  = {\bf (3^{(q_1,q_2)})^\star} + {\bf 6^\star} \;\;\; \mbox{and} \;\;\;
{\bf 3^{(p_{1,1},p_{2,1})}} \times {\bf (3^{(p_{1,2},p_{2,2})})^\star} = {\bf 1^{(q_1,q_2)}} + {\bf 8} \; ,
\ee
with $\mathrm{q}_1= \mathrm{p}_{1,1} + \mathrm{p}_{1,2} \, \mbox{mod} \, 2$ and $\mathrm{q}_2= \mathrm{p}_{2,1} + \mathrm{p}_{2,2} \, \mbox{mod} \, 2$, $\rm p_{i,j}=0,1$, 
especially we have
\be
{\bf 3^{(0,0)}} \times {\bf 3^{(0,0)}}  = {\bf (3^{(0,0)})^\star} + {\bf 6^\star} \;\;\; \mbox{and} \;\;\;
{\bf 3^{(0,0)}} \times {\bf (3^{(0,0)})^\star} = {\bf 1^{(0,0)}} + {\bf 8} \; .
\ee
All elements $g$, in the three-dimensional representations, can be uniquely written in the form
\be
g=\omega^o z^\zeta a^\alpha v^\chi x^\xi \;\;\; \mbox{with} \;\;\; o, \, \zeta, \, \alpha=0,1,2 \; , \; \chi=0,1,2,3 \; , \; \xi=0,1 \; .
\ee
Also here we display, for convenience, a representative (again in the three-dimensional representations) of each of the sixteen classes \cite{Sigma}
\bea
&&1 \, {\cal C}_1 \; : \; 1 \;\; , \;\;\; 9 \, {\cal C}_2 \; : \; v^2 \;\; , \;\;\;  1 \, {\cal C}_3 \; : \; a z a^2 z^2 =\omega \, 1 \;\; , \;\;\;  1 \, {\cal C}_3 \; : \; a^2 z a z^2 =\omega^2 \, 1 \;\; , \;\;\;
\\ \nonumber
&&  24 \, {\cal C}_3 \; : \; z \;\; , \;\;\; 18 \, {\cal C}_4 \; : \; v \;\; , \;\;\; 18 \, {\cal C}_4 \; : \; x \;\; , \;\;\;  18 \, {\cal C}_4 \; : \; \omega^2 \, v x \;\; , \;\;\; 9 \, {\cal C}_6 \; : \; \omega \, v^2 \;\; , \;\;\; 
\\ \nonumber
&& 9 \, {\cal C}_6 \; : \; \omega^2 \, v^2 \;\; , \;\;\; 18 \, {\cal C}_{12} \; : \; \omega \, v \;\; , \;\;\; 18 \, {\cal C}_{12} \; : \; \omega^2 \, v \;\; , \;\;\; 18 \, {\cal C}_{12} \; : \; \omega \, x \;\; , \;\;\; 18 \, {\cal C}_{12} \; : \; \omega^2 \, x \;\; , \;\;\; 
\\ \nonumber
&&18 \, {\cal C}_{12} \; : \; v x \;\; , \;\;\; 18 \, {\cal C}_{12} \; : \; \omega \,  v x \;\; .
\eea
Similarly to what we have seen for the representation ${\bf 3^{(0)}}$ of the group $\Sigma (36 \times 3)$ also here 30 elements of the group 
are represented by matrices with at least two degenerate eigenvalues and clearly these coincide with those already found for $\Sigma (36 \times 3)$.
The abelian subgroups of $\Sigma (72 \times 3)$ are: nine $Z_2$ symmetries, 13 $Z_3$ symmetries, 27 $Z_4$ groups, nine $Z_6$ symmetries, 27 $Z_{12}$ groups as well as four $Z_3 \times Z_3$ groups.
The $Z_2$ and $Z_6$ groups are all conjugate to each other, while the $Z_3$ symmetries form two categories (one with twelve members and the second one containing only the center of $SU(3)$).
The $Z_4$ groups and the $Z_{12}$ groups fall into three categories with nine members each which are conjugate to each other. In the following we will not use the four $Z_3 \times Z_3$ groups
as $G_e$ or $G_\nu$, since one of the $Z_3$ factors is the center of $SU(3)$. As one can see, also $\Sigma (72 \times 3)$ does not possess Klein subgroups.

Considering all possible admissible choices for the groups $G_e$ and $G_\nu$ we see that they have to be either both $Z_4$ subgroups or $Z_{12}$ subgroups of
$\Sigma (72 \times 3)$ or one is a $Z_4$, while the other one is a $Z_{12}$ subgroup. In all other cases the generators of $G_e$ and $G_\nu$ cannot give rise to the entire group $\Sigma (72 \times 3)$.
The mixing pattern arising from the admissible combinations is uniquely determined to be of the form
\small
\be
\label{eq:S72x3_1}
||U_{PMNS}||= \frac{1}{2 \sqrt{2}} \,\left(
\begin{array}{ccc}
\sqrt{3+\sqrt{3}} & \sqrt{4-\sqrt{3}} & 1\\
\sqrt{2} & \sqrt{3+\sqrt{3}} & \sqrt{3-\sqrt{3}}\\
\sqrt{3-\sqrt{3}} & 1 & \sqrt{4+\sqrt{3}}
\end{array}
\right)\approx
 \left(
\begin{array}{ccc}
0.769 & 0.532 & 0.354\\
0.500 & 0.769 & 0.398\\
0.398 & 0.354 & 0.846
\end{array}
\right) \; .
\ee
\normalsize
For the mixing angles follows $\sin^2 \theta_{12} \approx 0.324$, $\sin^2 \theta_{23} \approx 0.181$ and $\sin^2 \theta_{13} = 0.125$. Consequently, the value of $\chi^2$ is very large, $\chi^2 \approx 1984.1$, and only the
solar mixing angle turns out to be within the $3 \, \sigma$ range of the global fit \cite{global_latest}.
Similar results are found, if the second and third rows of the PMNS matrix in eq.(\ref{eq:S72x3_1}) are exchanged or the transpose (hermitian conjugate) of this matrix is considered as PMNS matrix. One interesting
feature of this mixing matrix is that the Jarlskog invariant does not vanish: $|J_{CP}|=\sqrt{3}/32 \approx 0.0541$. Examples of generators $g_e$ and $g_\nu$ of the subgroups $G_e$ and $G_\nu$ are:
for both $G_e$ and $G_\nu$ being $Z_4$ groups, $g_e=\omega^2 z^2 a^2 x$ and $g_\nu=z a^2 v$ can be chosen; for both $G_e$ and $G_\nu$ being $Z_{12}$ subgroups, we can take $g_e=z v x$
and $g_\nu=z^2 a v^2 x$, while for $G_e$ being a $Z_4$ and $G_\nu$ a $Z_{12}$ subgroup one possible combination of generators is $g_e=\omega^2 z^2 a^2 x$ and $g_\nu=z v x$.
If $G_e$ or $G_\nu$ is a $Z_2$ or a $Z_6$ subgroup, we can only fix one row or column of the PMNS matrix and the resulting mixing vectors coincide, as expected, with those found for the group $\Sigma (36 \times 3)$.
The observations which have been made for the matrices representing $Z_4$ and $Z_{12}$ and $Z_2$ and $Z_6$ generating elements, respectively, in $\Sigma (36 \times 3)$ are also valid in the case of the group
$\Sigma (72 \times 3)$ and thus clearly the pattern in eq.(\ref{eq:S72x3_1}) can be achieved with $Z_4$ or $Z_{12}$ subgroups and the mixing vectors arising from matrices representing $Z_2$ or $Z_6$ generating elements
have to be the same.

\mathversion{bold}
\subsection{Group $\Sigma (216 \times 3)$}
\mathversion{normal}
\label{sec33}

If we add the element $w$ which fulfills the relations \cite{Sigma}
\be
 w^9 = 1 \; , \;\; w^{-1} a w = z a \; , \;\; w^{-1} v^2 w = z v^2
\ee
to the generators $a$, $v$ and $z$ of the group $\Sigma (36 \times 3)$, we arrive at the group $\Sigma (216 \times 3)$ (identification number [[648, 532]] in the library SmallGroups).
Note that $x=w v w^{-1}$, see eqs.(\ref{eq:genrelS72x3},\ref{eq:genS72x3}), is also an element of this group.
For one of the faithful three-dimensional representations, denoted by ${\bf 3^{(0)}}$ in \cite{Sigma}, we choose the generators $a$, $v$, $z$ and $w$ to be represented by the matrices found in
eq.(\ref{eq:genS36x3}) and
\be
w= \left( \begin{array}{ccc}
 \epsilon & 0 & 0\\
 0 & \epsilon & 0\\
 0 & 0 & \epsilon \, \omega
\end{array}
\right) \;\;
\ee
with $\epsilon=e^{4 \pi i/9}$. Since $z w = w z$ holds, both generators $z$ and $w$ can be represented by diagonal matrices. Notice also that $w^3= \omega^2 1$ in the three-dimensional representation ${\bf 3^{(0)}}$.
The group $\Sigma (216 \times 3)$ possesses in total seven irreducible three-dimensional representations: one of them, ${\bf 3^{(a)}}$, is unfaithful  and real, while the six other ones are faithful
and form three pairs of complex conjugated representations ${\bf 3^{(p)}}$ and ${\bf (3^{(p)})^\star}$, $\rm p=0,1,2$. 
The representation matrices of the three faithful representations ${\bf 3^{(p)}}$ can be chosen as
\be
a \; , \;\; v \; , \;\; \omega^{\rm p} \, w \; , \;\; z  \;\;\; \mbox{with} \;\;\; \rm p=0,1,2 \; .
\ee
As is obvious from this, we can constrain ourselves to only consider the representation ${\bf 3^{(0)}}$  in our analysis of the mixing patterns. Relations among the different irreducible and faithful three-dimensional 
representations can also be found by exploiting the outer automorphisms of the group $\Sigma (216 \times 3)$, as we show in Appendix \ref{appS216x3}. 
Apart from the three-dimensional representations $\Sigma (216 \times 3)$ contains three one-, three two- and three eight-dimensional
representations, which are organized as one real representation and one complex conjugated pair. All these representations are unfaithful, while the three complex conjugated pairs of six-dimensional and the pair
of nine-dimensional representations are faithful in $\Sigma (216 \times 3)$. 

 The structure of the products of the faithful three-dimensional representations is similar to the one of the fundamental representation of $SU(3)$
 \be
{\bf 3^{(p_1)}} \times {\bf 3^{(p_2)}}  = {\bf (3^{(q_1)})^\star} + {\bf (6^{(q_1)})^\star} \;\;\; \mbox{and} \;\;\;
{\bf 3^{(p_1)}} \times {\bf (3^{(p_2)})^\star} = {\bf 1^{(q_2)}} + {\bf 8^{(q_2)}} 
\ee 
with $\mathrm{q}_1=-\mathrm{p}_1-\mathrm{p}_2 \, \mbox{mod} \, 3$, $\mathrm{q}_2=\mathrm{p}_1-\mathrm{p}_2 \, \mbox{mod} \, 3$ and $\rm p_{1,2}=0,1,2$. Especially, we have
\be
{\bf 3^{(0)}} \times {\bf 3^{(0)}}  = {\bf (3^{(0)})^\star} + {\bf (6^{(0)})^\star} \;\;\; \mbox{and} \;\;\;
{\bf 3^{(0)}} \times {\bf (3^{(0)})^\star} = {\bf 1^{(0)}} + {\bf 8^{(0)}} \; .
\ee
All elements $g$ of the group $\Sigma (216 \times 3)$ can be written, at least in the faithful three-dimensional representations, as
\be
g=\omega^o z^\zeta a^\alpha v^\chi x^\xi w^\kappa \;\;\; \mbox{with} \;\;\; o, \, \zeta, \, \alpha, \, \kappa=0,1,2 \; , \; \chi=0,1,2,3 \; , \; \xi=0,1 \; .
\ee
We mention one representative (in the faithful three-dimensional representations) for each of the 24 classes of this group \cite{Sigma}
\bea
&&1 \, {\cal C}_1 \; : \; 1 \;\; , \;\;\; 9 \, {\cal C}_2 \; : \; v^2 \;\; , \;\;\;  1 \, {\cal C}_3 \; : \; a z a^2 z^2 =\omega \, 1 \;\; , \;\;\;  1 \, {\cal C}_3 \; : \; a^2 z a z^2 =\omega^2 \, 1 \;\; , \;\;\; 
\\ \nonumber
&&24 \, {\cal C}_3 \; : \; z \;\; , \;\;\; 72 \, {\cal C}_3 \; : \; a w \;\; , \;\;\;  72 \, {\cal C}_3 \; : \; a w^2 \;\; , \;\;\;  54 \, {\cal C}_4 \; : \; v \;\; , \;\;\;  9 \, {\cal C}_6 \; : \; \omega \, v^2 \;\; , \;\;\; 
\\ \nonumber
&&9 \, {\cal C}_6 \; : \; \omega^2 \, v^2 \;\; , \;\;\;
12 \, {\cal C}_{9} \; : \; w \;\; , \;\;\; 12 \, {\cal C}_{9} \; : \; \omega \, w \;\; , \;\;\; 12 \, {\cal C}_{9} \; : \; \omega^2 \, w \;\; , \;\;\; 
\\ \nonumber
&& 12 \, {\cal C}_{9} \; : \; w^2 \;\; , \;\;\; 12 \, {\cal C}_{9} \; : \; \omega \, w^2 \;\; , \;\;\; 12 \, {\cal C}_{9} \; : \; \omega^2 \, w^2 \;\; , \;\;\; 54 \, {\cal C}_{12} \; : \; \omega \, v \;\; , \;\;\; 
\\ \nonumber
&&54 \, {\cal C}_{12} \; : \; \omega^2 \, v \;\; , \;\;\; 36 \, {\cal C}_{18} \; : \;  v^2 w \;\; , \;\;\; 36 \, {\cal C}_{18} \; : \;  \omega \, v^2 w \;\; , \;\;\; 36 \, {\cal C}_{18} \; : \; \omega^2 \,  v^2 w \;\; , \;\;\; 
\\ \nonumber
&&36 \, {\cal C}_{18} \; : \;  v w^2 \;\; , \;\;\; 36 \, {\cal C}_{18} \; : \;  \omega \, v w^2 \;\; , \;\;\; 36 \, {\cal C}_{18} \; : \; \omega^2 \,  v w^2  \;\; .
\eea
Most of the abelian subgroups of $\Sigma (216 \times 3)$ form one category in which all members are conjugate to each other: the nine $Z_2$ subgroups, the 27 $Z_4$ groups, the nine $Z_6$ groups, the twelve $Z_9$
symmetries, the 27 $Z_{12}$ groups as well as the 36 $Z_{18}$ symmetries. Only the 85 $Z_3$ groups form instead three categories of conjugate groups: one which only contains the center of $SU(3)$, one with twelve
members and the third one which comprises 72 $Z_3$ groups. Apart from these there are further abelian subgroups, $Z_3 \times Z_3$ and $Z_3 \times Z_9$, in which however one/the $Z_3$ factor is always given by the center
of $SU(3)$ so that these groups are not relevant for our discussion. Notice also the group $\Sigma (216 \times 3)$ does not possess Klein subgroups and as consequence, neutrinos
have to be either Dirac particles or $G_\nu$ can only be partly contained in $G_l=\Sigma (216 \times 3)$. 102 elements of this group are represented by matrices with at least two degenerate
eigenvalues in the representation ${\bf 3^{(0)}}$: these are the elements of the center of $SU(3)$, all $Z_2$ generating, all $Z_6$ as well as all $Z_9$ generating elements. All these cannot function as generators of the
groups $G_e$ and $G_\nu$; at least not alone, since the three generations of left-handed leptons cannot be distinguished in this way.

In table \ref{tab:Sigma216x3Pattern1} and \ref{tab:Sigma216x3Pattern2} on the next two pages we 
show the permutations of the mixing patterns with the smallest values of $\chi^2$ for all combinations of subgroups $G_e$ and $G_\nu$ which pass our constraints (i.e.
their generators give rise to the entire group $\Sigma (216 \times 3)$ 
and they allow to distinguish the three generations). 

\begin{landscape}
\begin{table}
\caption{\label{tab:Sigma216x3Pattern1} 
Mixing patterns associated with  
$\Sigma(216\times3)$. We mention the patterns, the subgroups $G_e$ and $G_\nu$, an example of their generators, the absolute value of $J_{CP}$,
the sine squares of the mixing angles as well as the value of $\chi^2$. In all cases $g_{e}$ and $g_{\nu}$
generate $\Sigma (216 \times 3)$. Note that we only give one example of $g_e$ and $g_\nu$. Analytic expressions of the
patterns are given if sufficiently simple. We use as figure of merit
$\chi^2$, computed using the results of the global fit in \cite{global_latest}
in table 1 in which the reactor fluxes are left free and short baseline reactor data are included (called ``Free
Fluxes + RSBL''). We always display the permutation of the pattern with the minimum value of $\chi^2$.
We show more than one permutation, if the $\chi^2$ value of the further pattern(s) is very close
to the minimum one. A star ($\star$) next to a result for a mixing angle $\sin^2 \theta_{ij}$ indicates
that its  contribution to the $\chi^2$ value is less than $9$ which is equivalent to being within the experimental $3 \, \sigma$ interval.
See the text for more details.}
\vspace{-0.2in}
\begin{center}
\renewcommand{\tabcolsep}{.9mm}
 \renewcommand{\arraystretch}{.9}
{\scriptsize
\begin{tabular}{|c|lccccc|}
\hline

&&&&&&\\
\# & Pattern & ($G_{e}, G_{\nu}$)  & Example of ($g_{e},
g_{\nu}$)&   $|J_{CP}|$  & ($\sin^2\theta_{12}, \sin^2\theta_{23}, \sin^2\theta_{13}$) & $\chi^2$ \\
&&&&&&\\
\hline 

&&&&&&\\
Ia &
$\frac{1}{2 \sqrt{2 (3+\sqrt{3})}} \; \left(
\begin{array}{ccc}
\sqrt{2} (2+\sqrt{3}) & \sqrt{2 (3+\sqrt{3})} & -1+\sqrt{3}\\
\sqrt{6} & \sqrt{2 (3+\sqrt{3})} & 3+\sqrt{3}\\
2 & 2 \sqrt{3+\sqrt{3}} & 2 \sqrt{2+\sqrt{3}}
\end{array}
\right)
\approx \left(
\begin{array}{ccc}
0.858     & 0.500    & 0.119 \\
 0.398 & 0.500   & 0.769 \\
0.325  & 0.707 & 0.628
\end{array}
\right)$
 & $\begin{array}{c}(Z_{18}, Z_{4})\\
(Z_{18}, Z_{12})
\end{array}$ & ($  \omega^{2} z a v x w,\, v $) &  0  &
$\begin{array}{l}
\sin^2\theta_{12}\approx0.254\\ \sin^2\theta_{23}=0.600\,  \star\\ \sin^2\theta_{13}\approx0.014\end{array}$ & 28.3 \\
&&&&&&\\

Ib &
$\frac{1}{2 \sqrt{2 (3+\sqrt{3})}} \; \left(
\begin{array}{ccc}
\sqrt{2} (2+\sqrt{3}) & \sqrt{2 (3+\sqrt{3})} & -1+\sqrt{3}\\
2 & 2 \sqrt{3+\sqrt{3}} & 2 \sqrt{2+\sqrt{3}}\\
\sqrt{6} & \sqrt{2 (3+\sqrt{3})} & 3+\sqrt{3}
\end{array}
\right)

\approx
\left(
\begin{array}{ccc}
 0.858 & 0.500 & 0.119 \\
 0.325 & 0.707 & 0.628 \\
 0.398 & 0.500 & 0.769
\end{array}
\right)
$
 & $\begin{array}{c}(Z_{18}, Z_{4})\\
(Z_{18}, Z_{12})
\end{array}$ & ($  \omega^{2} z a v x w,\, v $) &  0  &
$\begin{array}{l}
\sin^2\theta_{12}\approx0.254\\ \sin^2\theta_{23}=0.400\, \star\\ \sin^2\theta_{13}\approx 0.014\end{array}$ & 28.5 \\
&&&&&&\\

II &
$\frac{1}{2 \sqrt{3}} \; \left(
\begin{array}{ccc}
 3 & \sqrt{3} & 0\\
 \sqrt{2} & \sqrt{6} & 2\\
 1 & \sqrt{3} & 2 \sqrt{2}
\end{array}
\right)\approx \left(
\begin{array}{ccc}
 0.866 & 0.500 & 0 \\
 0.408 & 0.707 & 0.577\\
 0.289 & 0.500 & 0.816
\end{array}
\right)$
 & $\begin{array}{c}(Z_{18}, Z_{18})\\
\end{array}$ & ($ z a^{2} v^{2} x w,z v^{2} w $) &  0  &
$\begin{array}{l}
\sin^2\theta_{12}=0.250\\ \sin^2\theta_{23}\approx0.333 
\\ \sin^2\theta_{13}= 0\end{array}$ & 120.8 \\
&&&&&&\\

III & $
\left(
\begin{array}{ccc}
 0.844 & 0.525 & 0.110 \\
  0.449 & 0.804 & 0.390\\
 0.293 & 0.279 & 0.914
\end{array}
\right)$
 & $\begin{array}{c}(Z_{3}, Z_{18})\\
\end{array}$ &  ($\omega^{2} z^{2} a v^{3} x w,\, z^{2} v^{3} w $) &  0  &
$\begin{array}{l}
\sin^2\theta_{12}\approx0.279\,\star\\ \sin^2\theta_{23}\approx0.154\\ \sin^2\theta_{13}\approx0.012\end{array}$ & 131.2\\
&&&&&&\\

 IV &
 $\frac{1}{2 \sqrt{3+\sqrt{3}}} \,\left(
\begin{array}{ccc}
\sqrt{3 (3+\sqrt{3})} & \sqrt{3+\sqrt{3}} & 0\\
\sqrt{2 + \sqrt{3}} & \sqrt{3 (2+\sqrt{3})}& 2\\
1 & \sqrt{3} & 2 \sqrt{2+\sqrt{3}}
\end{array}
\right)\approx \left(
\begin{array}{ccc}
 0.866 & 0.500 & 0. \\
  0.444 & 0.769 & 0.460\\
 0.230 & 0.398 & 0.888
\end{array}
\right)$
 & $\begin{array}{c}(Z_{4}, Z_{18})\\
(Z_{12}, Z_{18})\\
\end{array}$ & ($ v,\,z^{2}  v^{2} w $)  &  0  &
$\begin{array}{l}
\sin^2\theta_{12}=0.250
\\ \sin^2\theta_{23}\approx0.211\\ \sin^2\theta_{13}= 0\end{array}$ & 175.7 \\
&&&&&&\\

V & $ \left(
\begin{array}{ccc}
  0.804   & 0.578   & 0.139 \\
 0.525   & 0.800 &  0.290  \\
 0.279   &   0.160 & 0.947
\end{array}
\right) $
 & $\begin{array}{c}(Z_{3}, Z_{4})\\
(Z_{3}, Z_{12})
\end{array}$ & ($\omega^{2} v w,\, v  $) &  0 &
$\begin{array}{l}
\sin^2\theta_{12}\approx0.341\\ \sin^2\theta_{23}\approx0.086\\ \sin^2\theta_{13}\approx 0.019 \,\star \end{array}$ & 182.7 \\
&&&&&&\\

\,\,\,\,VIa\,\,\,\,& $ \left(
\begin{array}{ccc}
 0.874 & 0.429 & 0.228 \\
 0.429 & 0.621 & 0.657 \\
 0.228 & 0.657 & 0.719
\end{array}
\right) $
 & $\begin{array}{c}(Z_{3}, Z_{3})\\
\end{array}$ &($\omega z^{2} a^{2}  x w, \omega^{2} z^{2} a v^{3} x w $) &  0.0321  &
$\begin{array}{l}
\sin^2\theta_{12}\approx0.194\\ \sin^2\theta_{23}\approx0.455\,\star\\ \sin^2\theta_{13}\approx 0.052\end{array}$ & 231.7 \\
&&&&&&\\

\,\,\,\,VIb\,\,\,\,& $ \left(
\begin{array}{ccc}
 0.874 & 0.429 & 0.228 \\
 0.228 & 0.657 & 0.719\\
 0.429 & 0.621 & 0.657 
\end{array}
\right) $
 & $\begin{array}{c}(Z_{3}, Z_{3})\\
\end{array}$ &($\omega z^{2} a^{2}  x w, \omega^{2} z^{2} a v^{3} x w $) &  0.0321  &
$\begin{array}{l}
\sin^2\theta_{12}\approx0.194\\ \sin^2\theta_{23}\approx0.545\,\star\\ \sin^2\theta_{13}\approx 0.052\end{array}$ & 235.5 \\
&&&&&&\\

\hline 
\end{tabular}}
\end{center}
\end{table}
\end{landscape}

\begin{landscape}
\begin{table}

\caption{\label{tab:Sigma216x3Pattern2} 
Mixing patterns associated with  
$\Sigma(216\times3)$. For further details see the caption of table \ref{tab:Sigma216x3Pattern1} and the text.}

\vspace{-10pt}
\begin{center}
\renewcommand{\tabcolsep}{.9mm}
 \renewcommand{\arraystretch}{.9}
{\scriptsize
\begin{tabular}{|c|lccccc|}
\hline

&&&&&&\\
\# & Pattern & ($G_{e}, G_{\nu}$)  & Example of ($g_{e},
g_{\nu}$)&   $|J_{CP}|$  & ($\sin^2\theta_{12}, \sin^2\theta_{23}, \sin^2\theta_{13}$) & $\chi^2$ \\
&&&&&&\\
\hline 

&&&&&&\\

VIIa &  $ 
\left(
\begin{array}{ccc}
 0.804 & 0.525 & 0.279 \\
 0.483 & 0.445 & 0.754 \\
 0.346 & 0.726 & 0.595
      \end{array}
\right) $
 & $\begin{array}{c}(Z_{4}, Z_{3})\\
(Z_{12}, Z_{3})
\end{array}$ &($v,\,\omega z^{2} a^{2}  x w $)  &  $\dfrac{1}{24}\approx0.0417$  &
$\begin{array}{l}
\sin^2\theta_{12}\approx0.299\, \star\\ \sin^2\theta_{23}\approx0.616\, \star\\ \sin^2\theta_{13}\approx 0.078\end{array}$ & 554.6 \\
&&&&&&\\

VIIb &  $ 
\left(
\begin{array}{ccc}
 0.804 & 0.525 & 0.279 \\
 0.346 & 0.726 & 0.595\\
0.483 & 0.445 & 0.754 
      \end{array}
\right) $
 & $\begin{array}{c}(Z_{4}, Z_{3})\\
(Z_{12}, Z_{3})
\end{array}$ &($v,\,\omega z^{2} a^{2}  x w $)  &  $\dfrac{1}{24}\approx0.0417$  &
$\begin{array}{l}
\sin^2\theta_{12}\approx0.299\, \star\\ \sin^2\theta_{23}\approx0.384\, \star\\ \sin^2\theta_{13}\approx 0.078\end{array}$ & 554.9 \\
&&&&&&\\

VIII & $   \left(
\begin{array}{ccc}
 0.804  &  0.525   & 0.279 \\
 0.517  & 0.723  &  0.458 \\
0.293 & 0.449   &  0.844
\end{array}
\right)$
 & $\begin{array}{c}(Z_{18}, Z_{3})\\
\end{array}$ & ($ z a^{2} v^{2} x w,\,\omega^{2} z^{2} a v^{3} x w $) &  $\dfrac{1}{12 \sqrt{3}}\approx0.0481$  &
$\begin{array}{l}
\sin^2\theta_{12}\approx0.299 \, \star\\ \sin^2\theta_{23}\approx0.227\\ \sin^2\theta_{13}\approx 0.078\end{array}$ & 608.7 \\
&&&&&&\\

IXa &
$ \frac{1}{2 \sqrt{6}}\left(
\begin{array}{ccc}
 \sqrt{3 (3+\sqrt{3})} & \sqrt{9 -\sqrt{3}} & \sqrt{2 (3-\sqrt{3})}\\
 \sqrt{3 (3-\sqrt{3})} & \sqrt{9+\sqrt{3}} & \sqrt{2 (3+\sqrt{3})}\\
 \sqrt{6} & \sqrt{6} & 2 \sqrt{3}
\end{array}
\right)  \approx
\left(
\begin{array}{ccc}
 0.769 & 0.550 & 0.325 \\
  0.398 & 0.669 & 0.628\\
 0.500 & 0.500 & 0.707
\end{array}
\right) $
 & $\begin{array}{c}(Z_{4}, Z_{18})\\
(Z_{12}, Z_{18})
\end{array}$ & ($v,\, z a^{2} v^{2} x w $) &  $\dfrac{1}{16}=0.0625$  &
$\begin{array}{l}
\sin^2\theta_{12}\approx0.339\, \star
\\ \sin^2\theta_{23}\approx0.441 \,\star\\ \sin^2\theta_{13}\approx 0.106\end{array}$ & 1255.5 \\
&&&&&&\\

IXb &
$ \frac{1}{2 \sqrt{6}}\left(
\begin{array}{ccc}
 \sqrt{3 (3+\sqrt{3})} & \sqrt{9 -\sqrt{3}} & \sqrt{2 (3-\sqrt{3})}\\
 \sqrt{6} & \sqrt{6} & 2 \sqrt{3}\\
 \sqrt{3 (3-\sqrt{3})} & \sqrt{9+\sqrt{3}} & \sqrt{2 (3+\sqrt{3})}
\end{array}
\right)  \approx
\left(
\begin{array}{ccc}
 0.769 & 0.550 & 0.325 \\
 0.500 & 0.500 & 0.707\\
 0.398 & 0.669 & 0.628
\end{array}
\right) $
 & $\begin{array}{c}(Z_{4}, Z_{18})\\
(Z_{12}, Z_{18})
\end{array}$ & ($v,\, z a^{2} v^{2} x w $) &  $\dfrac{1}{16}= 0.0625$  &
$\begin{array}{l}
\sin^2\theta_{12}\approx0.339\, \star
\\ \sin^2\theta_{23}\approx0.559 \,\star\\ \sin^2\theta_{13}\approx 0.106\end{array}$ & 1257.5 \\
&&&&&&\\

Xa &
$ \frac{1}{2 \sqrt{3}} \; \left(
\begin{array}{ccc}
 \sqrt{7} & \sqrt{3} & \sqrt{2}\\
 \sqrt{3} & \sqrt{3} & \sqrt{6}\\
 \sqrt{2} & \sqrt{6} & 2
\end{array}
\right) \approx \left(
\begin{array}{ccc}
 0.764 & 0.500 & 0.408 \\
 0.500 & 0.500 & 0.707 \\
 0.408 & 0.707 & 0.577
\end{array}
\right) $
 & $\begin{array}{c}(Z_{18},Z_{18})\\
\end{array}$ & ($ z a^{2} v^{2} x w,\,z a^{2} v^{3} w$) &  $\dfrac{1}{8\sqrt{3}}\approx 0.0722$ &
$\begin{array}{l}
\sin^2\theta_{12}=0.300 \,\star\\ \sin^2\theta_{23}=0.600\,\star\\ \sin^2\theta_{13}\approx 0.167\end{array}$ & 3753.2 \\

&&&&&&\\

Xb &
$ \frac{1}{2 \sqrt{3}} \; \left(
\begin{array}{ccc}
 \sqrt{7} & \sqrt{3} & \sqrt{2}\\
 \sqrt{2} & \sqrt{6} & 2\\
 \sqrt{3} & \sqrt{3} & \sqrt{6}
\end{array}
\right) \approx \left(
\begin{array}{ccc}
 0.764 & 0.500 & 0.408 \\
 0.408 & 0.707 & 0.577\\
 0.500 & 0.500 & 0.707 
\end{array}
\right) $
 & $\begin{array}{c}(Z_{18},Z_{18})\\
\end{array}$ & ($ z a^{2} v^{2} x w,\,z a^{2} v^{3} w$) &  $\dfrac{1}{8\sqrt{3}}\approx 0.0722$ &
$\begin{array}{l}
\sin^2\theta_{12}=0.300 \,\star\\ \sin^2\theta_{23}=0.400\,\star\\ \sin^2\theta_{13}\approx 0.167\end{array}$ & 3753.4 \\
&&&&&&\\
\hline 
\end{tabular}}
\end{center}
\end{table}
\end{landscape}

As one can see, only the two patterns Ia and Ib agree well with the experimental results for all mixing angles, so that they can be considered as a good leading order
form in a concrete model which only requires small corrections in order to achieve full accordance with the data. Three of the patterns lead to atmospheric and solar mixing angles which agree within $3 \, \sigma$ 
with the experimental data (i.e. both these angles carry a star ($\star$)), while for the reactor mixing angle there is only one such pattern, pattern V. Patterns with smaller values of $\chi^2$, i.e. $\chi^2 \lesssim 200$,
do not lead to a non-trivial Dirac phase, although all mixing angles are non-zero in the majority of the cases. On the other hand, patterns with larger $\chi^2$ in general give rise to non-vanishing $J_{CP}$ and thus
CP violation. Interestingly, almost all patterns VIIa through Xb in table \ref{tab:Sigma216x3Pattern2} accommodate both, solar and atmospheric, mixing angles well; however, the value 
of $\sin^2 \theta_{13}$ is too large (in all cases it is still at least a factor
of two smaller than the other two $\sin^2 \theta_{ij}$). Notice that we find cases in which pairs of $\{ G_e , G_\nu \}$ that are isomorphic, but not conjugate, do lead to different mixing patterns so that it is 
not sufficient to only specify the type of subgroups, e.g. the information $G_e=Z_4$ and $G_\nu=Z_{18}$ does not necessarily imply that we find the mixing pattern IV, because also the patterns IXa and IXb can arise from this combination of
subgroups. If we furthermore consider the combinations in which the roles of $G_e$ and $G_\nu$ are exchanged, the patterns Ia and Ib, see table \ref{tab:Sigma216x3Pattern1}, can result.

The pattern with the smallest value of $\chi^2$ ($\chi^2 \approx 10.6$) is obtained for $G_e=Z_3$ and $G_\nu=Z_{18}$. We do not list this case in table \ref{tab:Sigma216x3Pattern1} and \ref{tab:Sigma216x3Pattern2}, since
the generators of $G_e$ and $G_\nu$ do not give rise to the whole group $\Sigma (216 \times 3)$, but only to a group of order 162 with the mathematical structure $(Z_9 \times Z_3) \rtimes S_3$ \cite{struct16214}. It is
denoted by $[[162,14]]$ in the library SmallGroups. The mixing pattern reads
\small
\be
\label{S216x3_add}
||U_{PMNS}||= \frac{1}{\sqrt{6}} \; \left(
\begin{array}{ccc}
 2 c_{18} & \sqrt{2} & 2 s_{18}\\
  c_{18}-\sqrt{3} s_{18} & \sqrt{2} & \sqrt{3} c_{18} + s_{18}\\
 c_{18} + \sqrt{3} s_{18} & \sqrt{2} & \sqrt{3} c_{18} -s_{18}
\end{array}
\right)\approx
 \left(
\begin{array}{ccc}
 0.804 & 0.577 & 0.142 \\
 0.279 & 0.577 & 0.767 \\
 0.525 & 0.577 & 0.625
 \end{array}
\right)
\ee
\normalsize
where we have defined $s_{18}=\sin \pi/18 \approx 0.174$ and $c_{18}= \cos \pi/18 \approx 0.985$. The resulting mixing angles are: $\sin^2 \theta_{12} \approx 0.340$, $\sin^2 \theta_{23} \approx 0.601$
and $\sin^2 \theta_{13} \approx 0.020$ so that only the value of the solar mixing angle is marginally outside the $3 \, \sigma$ range \cite{global_latest}. 
The Jarlskog invariant vanishes and thus the Dirac phase is trivial. A fit with a value of $\chi^2$ only slightly larger than the optimal one, $\chi^2 \approx 10.8$,
can be achieved, if the second and third rows are exchanged in the PMNS matrix in eq.(\ref{S216x3_add}). Then the atmospheric mixing angle is determined to be $\sin^2 \theta_{23} \approx 0.399$ which is also well within the $3 \, \sigma$ 
range of \cite{global_latest}. Note that the PMNS matrix in eq.(\ref{S216x3_add}) can be written as the product of the TB mixing matrix and a rotation in the (13)-plane
by the angle $\theta=-\pi/18$: $||U_{PMNS}||=||U_{TB} R_{13} (-\pi/18)||$. If we exchange the second and third rows in eq.(\ref{S216x3_add}), $\theta$ has to change sign, $\theta=\pi/18$.\footnote{We use here
the same conventions as used in \cite{D96_D384,modular} for the signs in the TB mixing matrix $U_{TB}$ and for the definition of the angle $\theta$.} 
Such type of pattern has already been found \cite{D96_D384,modular,groupscan2,D6n2_others}, when studying groups belonging to the series $\Delta (6 n^2)$, $n \in \mathbb{N}$.
  In \cite{D96_D384,modular} it has been suggested that there might be a general rule, namely for $G_e=Z_3$, $G_\nu=Z_{2 n}$ and $G_l=\Delta (6 n^2)$ the angle $\theta$ reads
  $\theta=\pm\pi/(2 n)$. In our case one would have to choose $n=9$ and indeed the group [[162,14]] is a subgroup of $\Delta (6 n^2)$ for $n=9$.
  As example for the generators $g_e$ and $g_\nu$ of $G_e$ and $G_\nu$, respectively, we can take  $g_e=\omega^{2} z^{2} a v^{3} x w $ and $g_\nu= a^2 v x w $. 
  We note that the first column of the matrix in eq.(\ref{S216x3_add}) has been considered
as mixing vector in \cite{groupscan1}, while the authors of \cite{groupscan2} found the same mixing pattern in their scan for the group $(Z_{18} \times Z_6) \rtimes S_3$ (which has the identification number [[648,259]] in the library SmallGroups),
assuming that neutrinos are Majorana particles and $G_e$ is a $Z_3$ symmetry. Very recently, the same authors have performed a scan of smaller groups allowing neutrinos to be Dirac particles \cite{groupscan3} and
they also found the pattern in eq.(\ref{S216x3_add}) for the group with the identification number [[162,14]] in the library SmallGroups.

Assuming that one of the groups $G_e$ or $G_\nu$ is generated by an element with order 2 or 6 which is represented by a matrix with two degenerate eigenvalues in the representation ${\bf 3^{(0)}}$, we find 
as mixing vectors those which have already been found for the group $\Sigma (36 \times 3)$ and three additional ones that are: a vector which coincides  (in absolute values) with the second column of pattern Ia found in table \ref{tab:Sigma216x3Pattern1} and \ref{tab:Sigma216x3Pattern2},
a vector whose form is the one of the first row of pattern II and a vector with a form equal to the one of the second column of pattern III. The last mixing vector has also been considered in \cite{groupscan1} and has been found for the
groups with identification numbers [[162,14]] and [[486,61]] in the library SmallGroups, respectively. If we choose $G_e$ or $G_\nu$ to be generated by an element of order 9,
we find instead eight possible mixing vectors: a vector with only one non-vanishing entry and a vector with two equal entries (as regards the absolute value) and one vanishing one (these two are in common with 
the mixing vectors derived from $\Sigma (36 \times 3)$), a tri-maximal mixing vector, a vector with the same absolute values as the second and third rows of the TB mixing matrix, a vector which coincides (in absolute values)
with the third row of pattern Ia, a vector equal to the third column of pattern II, a vector of the same form as the first column of pattern III and a vector equal to the third column of pattern IV.
The vector coinciding with the first column of pattern III equals the columns, up to permutations, of a pattern mentioned in \cite{Delta27CP}, in which the mixing matrices arising from the flavor group $\Delta (27)$ and a CP
symmetry are analyzed.

As expected, we also find in our analysis of the patterns and mixing vectors of the group $\Sigma (216 \times 3)$ that $Z_4$ and $Z_{12}$ subgroups lead to the same patterns, if chosen as $G_e$ and/or $G_\nu$, and that the representation matrices of 
$Z_2$ and $Z_6$ generating elements are related via the matrices representing the non-trivial elements of the center of $SU(3)$.

\mathversion{bold}
\subsection{Group $\Sigma (360 \times 3)$}
\mathversion{normal}
\label{sec34}

The last group we discuss is $\Sigma (360 \times 3)$ which has 1080 elements. Its identification number in the library SmallGroups is [[1080,260]]. We can define it in terms of
four generators $a$, $f$, $h$ and $q$. The relations they fulfill can be found in \cite{Sigmanphi}
\bea
&&a^3=1 \; , \;\; f^2=1 \; , \;\; h^2=1 \; , \;\; q^2=1 \; ,
\\ \nonumber
&&(f q)^2=1 \; , \;\; (a h)^2=1 \; , \;\; (a f)^3=1 \; , \;\; (f h)^3=1\; , \;\;  (h q)^3=1  \; , \;\; (a q)^6=1 \; .
\eea
For an alternative set of generators (fulfilling a different set of relations) see also \cite{Sigmanphi}. We mention its character table in table \ref{tab:Sigma360x3_chartab}. This table is adapted from \cite{charsubsu3}.
This group has four irreducible faithful three-dimensional representations which form two complex conjugated pairs.
Apart from these representations $\Sigma (360 \times 3)$ has also a pair of complex conjugated
six- and fifteen-dimensional representations which are faithful. Out of the three nine-dimensional representations only two are faithful which are complex, while the unfaithful one is real. The other six unfaithful representations have dimension one, five (two representations), eight (two representations) and ten and are all real.

\begin{landscape}
\begin{table}
\caption[]{Character table of the group $\Sigma (360 \times 3)$, adapted from \cite{charsubsu3}. $c_1 \, {\cal C}_{c_2}$ denote the classes with $c_1$ elements which have order $c_2$.
$G$ is a representative of the class $c_1 \, {\cal C}_{c_2}$ in terms of the generators $a$, $f$, $h$ and $q$. Furthermore, $\rho=e^{2 \pi i/5}$ and $\zeta=e^{2 \pi i/15}$.
\label{tab:Sigma360x3_chartab}}

\vspace{-10pt}
\begin{center}
\scriptsize
\begin{tabular}{|c|ccccccccccccccccc|}
\hline
&\multicolumn{17}{|c|}{Classes}                                                 \\ \cline{2-18}
& $C_1$ &  $C_2$ &  $C_3$ &  $C_4$ &  $C_5$ &  $C_6$ &  $C_7$ &  $C_8$ &  $C_9$ &   $C_{10}$ & $C_{11}$  &$C_{12}$  &  $C_{13}$  & $C_{14}$  & $C_{15}$  & $C_{16}$  & $C_{17}$  \\
&$1 \, {\cal C}_1$& $45 \, {\cal C}_2$ &  $1 \, {\cal C}_3$ &  $1 \, {\cal C}_3$ & $120 \, {\cal C}_3$  & $120 \, {\cal C}_3$  &  $90 \, {\cal C}_4$ &  $72 \, {\cal C}_5$ & $72 \, {\cal C}_5$ & $45 \, {\cal C}_6$ & $45 \, {\cal C}_6$ &
$90 \, {\cal C}_{12}$ & $90 \, {\cal C}_{12}$ & $72 \, {\cal C}_{15}$ & $72 \, {\cal C}_{15}$  & $72 \, {\cal C}_{15}$  & $72 \, {\cal C}_{15}$ \\
\cline{1-18}
\rule[0.07in]{0cm}{0cm} $G$         &$1$ & $q$ & $(q a)^2$& $ (a q a)^2$ & $a$ & $h q a$ & $ f q h$ & $(h q a f q)^3$ & $ (h q a f q)^6$ & $q a$ & $q a^2$ & $a f q$ & $ a^2 f q$ & $ h q a f q $ & $(h q a f q)^2$ & $(h q a f q)^{13}$
& $(h q a f q)^{14}$\\
\cline{1-18}
\hline
${\bf 1}$                                          &1      &1      &1                            &1                               &1        &1      &1      &1                                &1                              &1                             &1               &1                 &1               &1                          &1   &1 &1          \\[0.1cm]
${\bf 3^{(1)}}$                                &3      &-1      &$3 \omega$        &$3 \omega^2 $      &0        &0      &1      &$-\rho^2-\rho^3$     &$-\rho-\rho^4$      &$-\omega^2$       &$-\omega$        &$\omega^2$        &$\omega$              &$-\zeta^2-\zeta^8$              &$-\zeta-\zeta^4$    &$-\zeta^{11}-\zeta^{14}$ &$-\zeta^7-\zeta^{13}$          \\[0.1cm]
${\bf 3^{(2)}}$                                &3      &-1      &$3 \omega$        &$3 \omega^2  $     &0        &0      &1      &$-\rho-\rho^4$         &$-\rho^2-\rho^3$   &$-\omega^2$       &$-\omega$        &$\omega^2$        &$\omega$          &$-\zeta^{11}-\zeta^{14}$      &$-\zeta^7-\zeta^{13}$    &$-\zeta^2-\zeta^8$ &$-\zeta-\zeta^4$          \\[0.1cm]
${\bf (3^{(1)})^\star}$                    &3      &-1      &$3 \omega^2$    &$3 \omega $          &0        &0      &1      &$-\rho^2-\rho^3$    &$-\rho-\rho^4$       &$-\omega$           &$-\omega^2$       &$\omega$        &$\omega^2$       &$-\zeta^7-\zeta^{13}$         &$-\zeta^{11}-\zeta^{14}$    &$-\zeta-\zeta^4$ &$-\zeta^2-\zeta^8$          \\[0.1cm]
${\bf (3^{(2)})^\star}$                    &3      &-1      &$3 \omega^2$    &$3 \omega $          &0        &0      &1      &$-\rho-\rho^4$        &$-\rho^2-\rho^3$    &$-\omega$           &$-\omega^2$      &$\omega$        &$\omega^2$       &$-\zeta-\zeta^4$            &$-\zeta^2-\zeta^8$    &$-\zeta^7-\zeta^{13}$  &$-\zeta^{11}-\zeta^{14}$          \\[0.1cm]
${\bf 5^{(1)}}$                                &5      &1      &5                             &5                              &2         &-1      &-1     &0                       &0                               &1                             &1                     &-1             &-1                &0                      &0     &0      &0      \\[0.1cm]
${\bf 5^{(2)}}$                                &5      &1      &5                             &5                              &-1        &2      &-1      &0                      &0                                &1                           &1                 &-1                   &-1                &0                      &0        &0      &0     \\[0.1cm]
${\bf 6}$                                          &6      &2      &$6 \omega $         &$6 \omega^2 $    &0          &0      &0      &1                         &1                                &$2 \omega^2$   &$2 \omega$        &0                 &0              &$\omega$              &$\omega^2$     &$\omega$      &$\omega^2$       \\   [0.1cm]
${\bf 6^\star}$                                &6      &2      &$6 \omega^2 $     &$6 \omega $         &0         &0      &0      &1                          &1                               &$2 \omega$       &$2 \omega^2$      &0               &0              &$\omega^2$                    &$\omega$     &$\omega^2$      &$\omega$      \\[0.1cm]
${\bf 8^{(1)}}$                                &8      &0      &8                             &8                              &-1        &-1      &0     &$-\rho^2-\rho^3$   &$-\rho-\rho^4$        &0                           &0                &0                      &0                &$-\rho-\rho^4$          &$-\rho^2-\rho^3$   &$-\rho^2-\rho^3$ &$-\rho-\rho^4$          \\[0.1cm]
${\bf 8^{(2)}}$                                &8      &0      &8                            &8                                &-1       &-1      &0     &$-\rho-\rho^4$       &$-\rho^2-\rho^3$     &0                          &0                 &0                 &0                 &$-\rho^2-\rho^3$              &$-\rho-\rho^4$   &$-\rho-\rho^4$ &$-\rho^2-\rho^3$          \\[0.1cm]
${\bf 9^{(1)}}$                                &9      &1      &9                            &9                               &0        &0        &1      &-1                              &-1                    &1                &1                  &1              &1            &-1                     &-1     &-1       &-1      \\[0.1cm]
${\bf 9^{(2)}}$                                &9      &1      &$9 \omega $        &$9 \omega^2$      &0        &0         &1      &-1                              &-1                      &$\omega^2$         &$\omega$          &$\omega^2$   &$\omega$        &$-\omega$             &$-\omega^2$   &$-\omega$ &$-\omega^2$          \\[0.1cm]
${\bf (9^{(2)})^\star}$                  &9      &1      &$9 \omega^2 $     &$9 \omega $         &0       &0          &1      &-1                               &-1                      &$\omega$           &$\omega^2$         &$\omega$       &$\omega^2$    &$-\omega^2$            &$-\omega$     &$-\omega^2$       &$-\omega$       \\[0.1cm]
${\bf 10}$                                   &10      &-2      &10                         &10                            &1          &1      &0        &0                                &0                 &-2                &-2               &0                &0            &0                         &0       &0       &0       \\[0.1cm]
${\bf 15}$                                    &15    &-1      &$15 \omega $           &$15 \omega^2$   &0        &0          &-1      &0                               &0                    &$-\omega^2$      &$-\omega$         &$-\omega^2$       &$-\omega$     &0                   &0      &0       &0       \\[0.1cm]
${\bf 15^\star}$                          &15      &-1      &$15 \omega^2 $    &$15 \omega $        &0        &0         &-1      &0                                &0                    &$-\omega$         &$-\omega^2$        &$-\omega$       &$-\omega^2$    &0                    &0    &0       &0       \\
\hline
\end{tabular}
\end{center}

\end{table}
\end{landscape}

\normalsize
The form of the product of ${\bf 3^{(1)}}$ with itself and with its complex conjugated representation is like in $SU(3)$  \footnote{The form of the products is very similar for ${\bf 3^{(2)}}$. The combinations ${\bf 3^{(1)}} \times {\bf 3^{(2)}}$
and ${\bf 3^{(1)}} \times ({\bf 3^{(2)}})^\star$ and alike are instead irreducible in $\Sigma (360 \times 3)$.}
\be
{\bf 3^{(1)}} \times {\bf 3^{(1)}}  = {\bf (3^{(1)})^\star} + {\bf 6^\star} \;\;\; \mbox{and} \;\;\;
{\bf 3^{(1)}} \times {\bf (3^{(1)})^\star} = {\bf 1} + {\bf 8^{(1)}} \; .
\ee
 The generators of the group can be represented by the following matrices in the representation ${\bf 3^{(1)}}$:  $a$ can be chosen as in eq.(\ref{eq:genS36x3}) and
\be
\label{eq:genS360x3}
f= \left( \begin{array}{ccc}
 1 & 0 & 0\\
 0 & -1 & 0\\
 0 & 0 & -1
 \end{array}
\right) \;\; , \;\;\;
h= \frac{1}{2} \, \left( \begin{array}{ccc}
 -1 & \mu_{-} & \mu_{+} \\
 \mu_{-}  & \mu_{+}  & -1\\
 \mu_{+}  & -1 & \mu_{-}
 \end{array}
\right) \;\; , \;\;\;
q= \left( \begin{array}{ccc}
 -1 & 0 & 0\\
 0 & 0 & -\omega\\
 0 & -\omega^2 & 0
 \end{array}
\right) 
\ee
with $\mu_{\pm}=\frac{1}{2} \left(-1 \pm \sqrt{5} \right)$. All these matrices have determinant +1. A set of representation matrices for the other three-dimensional representations ${\bf 3^{(2)}}$, ${\bf( 3^{(1)})^\star}$ and 
${\bf( 3^{(2)})^\star}$ can be constructed by using outer automorphisms of the group $\Sigma (360 \times 3)$ which relate the other representations and their matrix realizations to ${\bf 3^{(1)}}$ and $a$, $f$, $h$ and $q$, see 
eqs. (\ref{eq:genS36x3},\ref{eq:genS360x3}),  
\bea
{\bf 3^{(1)}} \rightarrow {\bf 3^{(2)}} &:& \;\; a \; \rightarrow \; a h q  \;\; , \;\; f \; \rightarrow \; f a h f  \;\; , \;\; h \; \rightarrow \; q \;\; , \;\; q \; \rightarrow \; a^2 h  \;\; , \;\;
\\ \nonumber
{\bf 3^{(1)}} \rightarrow {\bf (3^{(1)})^\star} &:& \;\; a \; \rightarrow \; a^2 h q  \;\; , \;\; f \; \rightarrow \; f a^2 h f  \;\; , \;\; h \; \rightarrow \; q \;\; , \;\; q \; \rightarrow \; a h  \;\; , \;\;
\\ \nonumber
{\bf 3^{(1)}} \rightarrow {\bf (3^{(2)})^\star} &:& \;\; a \; \rightarrow \; a^2  \;\; , \;\; f \; \rightarrow \; f  \;\; , \;\; h \; \rightarrow \; h \;\; , \;\; q \; \rightarrow \; q  \;\; .
\eea
These outer automorphisms can be nicely seen as symmetries of the character table: the first one exchanges the representations ${\bf 3^{(1)}}$ and ${\bf 3^{(2)}}$ as well as their
 complex conjugates and ${\bf 8^{(1)}}$ with  ${\bf 8^{(2)}}$ and at the same time the classes: $C_8 \leftrightarrow C_9$, $C_{14} \leftrightarrow C_{16}$ and $C_{15} \leftrightarrow C_{17}$,
while the second one exchanges all representations with their complex conjugates, i.e. ${\bf 3^{(1)}}$ with ${\bf (3 ^{(1)})^\star}$, ${\bf 3^{(2)}}$ with ${\bf (3 ^{(2)})^\star}$, ${\bf 6}$ and ${\bf 6}^\star$,
${\bf 9^{(2)}}$ with ${\bf (9 ^{(2)})^\star}$ and ${\bf 15}$ with ${\bf 15}^\star$ and it also exchanges the classes: $C_3 \leftrightarrow C_4$, $C_{10} \leftrightarrow C_{11}$, $C_{12} \leftrightarrow C_{13}$,
$C_{14} \leftrightarrow C_{17}$ and $C_{15} \leftrightarrow C_{16}$.
The third automorphism exchanges the representations  ${\bf 3^{(1)}}$ and ${\bf (3^{(2)})^\star}$ as well as their complex conjugates, ${\bf 6}$ with ${\bf 6^\star}$, the two
real eight-dimensional representations ${\bf 8^{(1)}}$ and ${\bf 8^{(2)}}$, the complex conjugated pair of nine-dimensional representations as well as the fifteen-dimensional representations; at the same
time the following classes have to be exchanged: $C_3 \leftrightarrow C_4$,  $C_8 \leftrightarrow C_9$,   $C_{10} \leftrightarrow C_{11}$,  $C_{12} \leftrightarrow C_{13}$,  $C_{14} \leftrightarrow C_{15}$ and
 $C_{16} \leftrightarrow C_{17}$.  The group of outer automorphisms is a Klein group $Z_2 \times Z_2$. Thus, the mentioned mappings have order two and commute among each other, as can be checked.

The abelian subgroups of $\Sigma (360 \times 3)$ which are generated by a single element are 45 $Z_2$ groups, 121 $Z_3$ symmetries, 45 $Z_4$ symmetries, 36 $Z_5$ groups, 45 $Z_6$ symmetries, 
45 $Z_{12}$ groups as well as 36 $Z_{15}$ symmetries.
All these $Z_n$ subgroups are conjugate among each other apart from the $Z_3$ symmetries which form three categories (one of which contains only the center of $SU(3)$, a second one with 60 conjugate members
as well as a third one with also 60 members). There are 30 Klein groups which form two categories comprising 15 conjugate groups each. There are also abelian subgroups of the form
$Z_3 \times Z_3$ and $Z_2 \times Z_6$ which however contain the center of $SU(3)$ and thus are not relevant for our discussion.

The patterns we find for $\Sigma (360 \times 3)$ are listed in table \ref{tab:Sigma360x3Pattern1}-\ref{tab:Sigma360x3Pattern3}. Patterns Ia through XVIII all arise from subgroups $G_e$ and $G_\nu$ which are
generated by a single element each and thus require neutrinos to be Dirac particles. The only exception are the patterns XIa and XIb which can also be achieved, if both groups $G_e$ and $G_\nu$ are Klein
symmetries. Furthermore, there are several patterns, called KI through KVII, that are generated, if one of the two residual symmetries, $G_e$ or $G_\nu$, is a Klein group. Indeed for all patterns apart from KI the permutation with
the minimum value of $\chi^2$ is compatible with $G_\nu$ being a Klein group. Thus, neutrinos can naturally be Majorana particles. Overall there are only a few patterns with a value of $\chi^2$ smaller than 100: patterns
Ia through IIIb. Unfortunately, all these predict vanishing CP violation. This is very similar to what we have already observed in the case of the group $\Sigma (216 \times 3)$, see table \ref{tab:Sigma216x3Pattern1} 
and \ref{tab:Sigma216x3Pattern2}. The pattern with the smallest $\chi^2$ ($\chi^2 \approx 338.0$) and with non-vanishing $J_{CP}$ is pattern VII which, however, does not fit any of the three mixing angles at the $3 \, \sigma$
level or better. Pattern XVIII turns out to be the only one which leads to maximal CP violation, $|\sin \delta|=1$, as we indicate with a circle ($\circ$) in table \ref{tab:Sigma360x3Pattern3}. Although CP violation is not maximal
in the other cases, we frequently find patterns with $|\sin \delta| \gtrsim 0.9$. 
Almost half of the patterns Ia through XVIII fits the atmospheric mixing angle well, while none leads to a reactor mixing angle which is compatible at the $3 \, \sigma$ level with the results found in \cite{global_latest}.
On top of that only pattern Ia is able to accommodate two of the three mixing angles well. Among the patterns Ki, i=I, ..., VII, the smallest value of $\chi^2$ is $\chi^2 \approx 148.1$ which is larger than that of the patterns
Ia to IIIb. Although pattern KI has $\chi^2 \approx 148.1$ and only fits the solar mixing angle well, it might be interesting, since $J_{CP}$ is non-vanishing in this case. This is at variance to what we have found for the patterns
which are produced in setups with $G_e$ and $G_\nu$ being both generated by one element each. It is likely to be a coincidence, but indeed all patterns Ki, i=I, ..., VII lead to non-zero $J_{CP}$. Two of these, 
pattern KV and KVI, have also been mentioned in \cite{Hu_paper}.
Generally speaking, all patterns would require additional ingredients to be present which allow them to be compatible with the experimental data \cite{global_latest}. These might be suitable corrections coming from
contributions to the mass matrices of charged leptons and neutrinos which are not invariant under $G_e$ and $G_\nu$, respectively,
 in a concrete model or modifications of the breaking pattern, e.g. by involving CP \cite{flavorCP_1,flavorCP_2}.
As one can see from tables \ref{tab:Sigma360x3Pattern1}-\ref{tab:Sigma360x3Pattern3}, the choice of $Z_4$ or $Z_{12}$ subgroups and of $Z_5$ or $Z_{15}$ subgroups, respectively, leads to the same mixing patterns.
This is, like in the other groups $\Sigma (n \varphi)$ with $\varphi=3$, due to the relation of these subgroups via the center of $SU(3)$.

As one can check there are 138 matrices among the representation matrices in ${\bf 3^{(1)}}$ which have at least two degenerate eigenvalues: these are all $Z_2$ and all $Z_6$ generating elements as well as
the elements which comprise the center of $SU(3)$. The former can be identified with the generators of $G_e$ or $G_\nu$ and then give rise to one of the fourteen different mixing vectors: the three vectors which have been found in the 
analysis of the group $\Sigma (36 \times 3)$, a vector of the same form (in absolute values) as the first column of the TB mixing matrix, a vector which is tri-maximal, a vector which equals
 the first row of pattern KI (see table \ref{tab:Sigma360x3Pattern3}), another one of the same form of the third row of pattern KI, a vector equal to the second column of pattern KII, 
 another one which is of the same form as the third column of pattern KII, a vector which
 coincides with the third column of pattern KIII, vectors like the second and the third columns of pattern KIVa, a vector with the same form as the third column of pattern KV as well as a vector which coincides with the second
 column of pattern KVI. 
  

\begin{landscape}
\begin{table}
\caption{\label{tab:Sigma360x3Pattern1} Mixing patterns associated with 
$\Sigma(360\times3)$. 
For more details see the caption of table \ref{tab:Sigma216x3Pattern1} and the text.}
\vspace{-20pt}
\begin{center}
\renewcommand{\tabcolsep}{.1mm}
 \renewcommand{\arraystretch}{.7}
{\scriptsize
\tiny
\begin{tabular}{|c|lccccc|}
\hline

&&&&&&\\
\# & Pattern & ($G_{e}, G_{\nu}$)  & Example of ($g_{e},
g_{\nu}$)&   $|J_{CP}|$  & ($\sin^2\theta_{12}, \sin^2\theta_{23}, \sin^2\theta_{13}$) & $\quad\chi^2\quad$ \\
&&&&&&\\
\hline 

&&&&&&\\

Ia & $
\frac 14 \; \left(
\begin{array}{ccc}
1+\sqrt{5} & \sqrt{5-\sqrt{3}-\sqrt{5}+\sqrt{15}} & \sqrt{5+\sqrt{3}-\sqrt{5}-\sqrt{15}}\\
\sqrt{5-\sqrt{3}-\sqrt{5}+\sqrt{15}}  & \sqrt{8+\sqrt{3}-\sqrt{15}} & \sqrt{3 + \sqrt{5}}\\
\sqrt{5+\sqrt{3}-\sqrt{5}-\sqrt{15}} & \sqrt{3 + \sqrt{5}}\ & \sqrt{8-\sqrt{3}+\sqrt{15}}
\end{array}
\right)
\approx
\left(
\begin{array}{ccc}
 0.809 & 0.554 & 0.197 \\
 0.554 & 0.605 & 0.572\\
 0.197 & 0.572 & 0.796
\end{array}
\right)$& $\begin{array}{c}(Z_{4},Z_{4})\\(Z_{4},Z_{12})\\(Z_{12},Z_{12})
\end{array}$  &
($f q h$, $a f h q a^2$)
& 0 & $\begin{array}{l}
\sin^2\theta_{12}\approx0.319\,\star \\ \sin^2\theta_{23}\approx0.341\,\star\\ \sin^2\theta_{13}\approx 0.039\end{array}$ & 58.0\\
&&&&&&\\

Ib & $
\frac 14 \; \left(
\begin{array}{ccc}
1+\sqrt{5} & \sqrt{5-\sqrt{3}-\sqrt{5}+\sqrt{15}} & \sqrt{5+\sqrt{3}-\sqrt{5}-\sqrt{15}}\\
\sqrt{5+\sqrt{3}-\sqrt{5}-\sqrt{15}} & \sqrt{3 + \sqrt{5}}\ & \sqrt{8-\sqrt{3}+\sqrt{15}}\\
\sqrt{5-\sqrt{3}-\sqrt{5}+\sqrt{15}}  & \sqrt{8+\sqrt{3}-\sqrt{15}} & \sqrt{3 + \sqrt{5}}
\end{array}
\right)
\approx
\left(
\begin{array}{ccc}
 0.809 & 0.554 & 0.197 \\
 0.197 & 0.572 & 0.796 \\
 0.554 & 0.605 & 0.572
\end{array}
\right)$& $\begin{array}{c}(Z_{4},Z_{4})\\(Z_{4},Z_{12})\\(Z_{12},Z_{12})
\end{array}$  &
($f q h$, $a f h q a^2$)
& 0 & $\begin{array}{l}
\sin^2\theta_{12}\approx0.319\, \star \\ \sin^2\theta_{23}\approx0.659\\ \sin^2\theta_{13}\approx 0.039\end{array}$ &58.9\\
&&&&&&\\

IIa & $
\frac{1}{20} \;
\left(
\begin{array}{ccc}
 5-\sqrt{5}+\sqrt{30 (5+\sqrt{5})} & 4 \sqrt{5} & 2 \sqrt{5 (7-\sqrt{5} -\sqrt{6 (5-\sqrt{5})})}\\
4 \sqrt{5} & -5+\sqrt{5}+\sqrt{30 (5+\sqrt{5})}  &2  \sqrt{5 (7-\sqrt{5} +\sqrt{6 (5-\sqrt{5})})}\\
2 \sqrt{5 (7-\sqrt{5} -\sqrt{6 (5-\sqrt{5})})} & 2  \sqrt{5 (7-\sqrt{5} +\sqrt{6 (5-\sqrt{5})})} & 2 (5+\sqrt{5})
\end{array}
\right)
\approx
\left(
\begin{array}{ccc}
 0.875 & 0.447 & 0.186 \\
 0.447 & 0.598 & 0.665 \\
 0.186 & 0.665 & 0.724
\end{array}
\right)$& $\begin{array}{c}(Z_{5},Z_{5})\\(Z_{5},Z_{15})\\(Z_{15},Z_{15})
\end{array}$ &
($( h q a f q)^3$, $a^2 f q h$)
& 0 & $\begin{array}{l}
\sin^2\theta_{12}\approx0.207\\ \sin^2\theta_{23}\approx0.458\, \star\\ \sin^2\theta_{13}\approx 0.035\end{array}$   & 84.6\\
&&&&&&\\

IIb & $
\frac{1}{20} \;
\left(
\begin{array}{ccc}
 5-\sqrt{5}+\sqrt{30 (5+\sqrt{5})} & 4 \sqrt{5} & 2 \sqrt{5 (7-\sqrt{5} -\sqrt{6 (5-\sqrt{5})})}\\
 2 \sqrt{5 (7-\sqrt{5} -\sqrt{6 (5-\sqrt{5})})} & 2  \sqrt{5 (7-\sqrt{5} +\sqrt{6 (5-\sqrt{5})})} & 2 (5+\sqrt{5})\\
4 \sqrt{5} & -5+\sqrt{5}+\sqrt{30 (5+\sqrt{5})}  &2  \sqrt{5 (7-\sqrt{5} +\sqrt{6 (5-\sqrt{5})})}
\end{array}
\right)
\approx
\left(
\begin{array}{ccc}
 0.875 & 0.447 & 0.186 \\
 0.186 & 0.665 & 0.724 \\
 0.447 & 0.598 & 0.665
\end{array}
\right)$& $\begin{array}{c}(Z_{5},Z_{5})\\(Z_{5},Z_{15})\\(Z_{15},Z_{15})
\end{array}$ &
($( h q a f q)^3$, $a^2 f q h$)
&  0 & $\begin{array}{l}
\sin^2\theta_{12}\approx0.207\\ \sin^2\theta_{23}\approx0.542\, \star\\ \sin^2\theta_{13}\approx 0.035\end{array}$  & 88.9\\
&&&&&&\\

IIIa & $
\left(
\begin{array}{ccc}
 0.894 & 0.430 & 0.124 \\
 0.250 & 0.710 & 0.659 \\
 0.372 & 0.558 & 0.742
\end{array}
\right)$& $\begin{array}{c}(Z_{5},Z_{4})\\(Z_{15},Z_{4})\\(Z_{5},Z_{12})\\(Z_{15},Z_{12})
\end{array}$   &
($(h q a f q)^3$, $a^2 f h q a$)
& 0 &$\begin{array}{l}
\sin^2\theta_{12}\approx0.188\\ \sin^2\theta_{23}\approx0.441 \, \star\\ \sin^2\theta_{13}\approx 0. 015\end{array}$ &93.0\\
&&&&&&\\

IIIb & $
\left(
\begin{array}{ccc}
 0.894 & 0.430 & 0.124 \\
 0.372 & 0.558 & 0.742\\
  0.250 & 0.710 & 0.659
\end{array}
\right)$&
$\begin{array}{c}(Z_{5},Z_{4})\\(Z_{15},Z_{4})\\(Z_{5},Z_{12})\\(Z_{15},Z_{12})
\end{array}$   &
($(h q a f q)^3$, $a^2 f h q a$) & 0 &$\begin{array}{l}
\sin^2\theta_{12}\approx0.188\\ \sin^2\theta_{23}\approx0.559 \, \star\\ \sin^2\theta_{13}\approx 0. 015\end{array}$ &95.0\\
&&&&&&\\

IV & $
\frac{1}{20} \;
\left(
\begin{array}{ccc}
2 \sqrt{5 (7+\sqrt{5} + \sqrt{6 (5+\sqrt{5})})} & 4 \sqrt{5} & -5-\sqrt{5}+\sqrt{30 (5-\sqrt{5})} \\
2 \sqrt{5 (7+\sqrt{5} - \sqrt{6 (5+\sqrt{5})})} & 5+\sqrt{5}+\sqrt{30 (5-\sqrt{5})} & 4 \sqrt{5}\\
2 (5-\sqrt{5}) & 2 \sqrt{5 (7+\sqrt{5} - \sqrt{6 (5+\sqrt{5})})}  & 2 \sqrt{5 (7+\sqrt{5} + \sqrt{6 (5+\sqrt{5})})}
\end{array}
\right)
\approx
\left(
\begin{array}{ccc}
 0.890 & 0.447 & 0.093 \\
 0.364 & 0.817 & 0.447 \\
 0.276 & 0.364 & 0.890
\end{array}
\right)$ & $\begin{array}{c}(Z_{5},Z_{5})\\(Z_{5},Z_{15})\\(Z_{15},Z_{15})
\end{array}$  &
($(h q a f q)^3$, $q h q a f h$)
& 0 & $\begin{array}{l}
\sin^2\theta_{12}\approx0.202\\ \sin^2\theta_{23}\approx0.202\\ \sin^2\theta_{13}\approx 0.009 \,  \end{array}$&171.0\\
&&&&&&\\

V & $
\left(
\begin{array}{ccc}
 0.795 & 0.602 & 0.076 \\
 0.590 & 0.739 & 0.324 \\
 0.139 & 0.302 & 0.943
\end{array}
\right)$& $\begin{array}{c}(Z_{5},Z_{4})\\(Z_{15},Z_{4})\\(Z_{5},Z_{12})\\(Z_{15},Z_{12})
\end{array}$   &
($(h q a f q)^3$, $f q h$)
& 0 &  $\begin{array}{l}
\sin^2\theta_{12}\approx0.364\\ \sin^2\theta_{23}\approx 0.106\\ \sin^2\theta_{13}\approx 0.006 \,\end{array}$&227.7\\
&&&&&&\\

 VI & $
\frac{1}{2 \sqrt{5}} \;
 \left(
\begin{array}{ccc}
\sqrt{2 (5+\sqrt{15})} & \sqrt{2 (5-\sqrt{15})} & 0\\
\sqrt{5-\sqrt{15}} & \sqrt{5+\sqrt{15}} & \sqrt{10}\\
\sqrt{5-\sqrt{15}} & \sqrt{5+\sqrt{15}} & \sqrt{10}
\end{array}
\right)
\approx
 \left(
\begin{array}{ccc}
 0.942 & 0.336 & 0 \\
 0.237 & 0.666 & 0.707 \\
 0.237 & 0.666 & 0.707
\end{array}
\right)$ & $\begin{array}{c}(Z_{5},Z_{4})\\(Z_{15},Z_{4})\\(Z_{5},Z_{12})\\(Z_{15},Z_{12})
\end{array}$  &
($(h q a f q)^3$, $q a f a^2$)
& 0 & $\begin{array}{l}
\sin^2\theta_{12}\approx0.113\\ \sin^2\theta_{23}=0.500\,\star\\ \sin^2\theta_{13}= 0 \end{array}$ & 328.2\\

&&&&&&\\

VII & $
\frac 16 \;
\left(
\begin{array}{ccc}
2 \sqrt{3+\sqrt{5}} & \sqrt{2 (9-\sqrt{5})} & -1+\sqrt{5}\\
 -1+\sqrt{5} &  -1+\sqrt{5} & 2 \sqrt{6+\sqrt{5}}\\
\sqrt{2 (9-\sqrt{5})}  & 2 \sqrt{3+\sqrt{5}} & -1+\sqrt{5}
\end{array}
\right)
\approx
\left(
\begin{array}{ccc}
 0.763 & 0.613 & 0.206 \\
 0.206 & 0.206 & 0.957\\
 0.613 & 0.763 & 0.206
\end{array}
\right)$& $\begin{array}{c}(Z_{3},Z_{3})
\end{array}$  & 
($h q a$, $h f$)
& $\begin{array}{c}
\frac{1}{108} \sqrt{6 (3-\sqrt{5})} \\
\approx 0.0198
\end{array}$
&  $\begin{array}{l}
\sin^2\theta_{12}\approx 0.392\, \\ \sin^2\theta_{23}\approx0.956\\ \sin^2\theta_{13}\approx 0.042\end{array}$ & 338.0\\
&&&&&&\\

\hline
\end{tabular}}
\end{center}
\end{table}
\end{landscape}

\vspace{-20pt}
\begin{landscape}

\begin{table}
\caption{\label{tab:Sigma360x3Pattern2} Mixing patterns associated with  
$\Sigma(360\times3)$. Note that patterns XIa and XIb can also be achieved, if both $G_e$ and $G_\nu$
are Klein subgroups of $\Sigma (360 \times 3)$.
For further explanation see the caption of table \ref{tab:Sigma360x3Pattern1}.}
\vspace{-20pt}
\begin{center}
\renewcommand{\tabcolsep}{.1mm}
 \renewcommand{\arraystretch}{.7}
{\scriptsize
\tiny
\begin{tabular}{|c|lccccc|}
\hline

&&&&&&\\
\# & Pattern & ($G_{e}, G_{\nu}$)  & Example of ($g_{e},
g_{\nu}$)&   $|J_{CP}|$  & ($\sin^2\theta_{12}, \sin^2\theta_{23}, \sin^2\theta_{13}$) & $\quad\chi^2\quad$ \\
&&&&&&\\
\hline 

&&&&&&\\

VIII & $ \frac 14 \; \left(
\begin{array}{ccc}
\sqrt{5+\sqrt{3}+\sqrt{5}+\sqrt{15}} & \sqrt{8-\sqrt{6 (3+\sqrt{5})}} & \sqrt{3-\sqrt{5}}\\
-1+\sqrt{5} &  \sqrt{5+\sqrt{3}+\sqrt{5}+\sqrt{15}} & \sqrt{5-\sqrt{3}+\sqrt{5}-\sqrt{15}}\\
\sqrt{5-\sqrt{3}+\sqrt{5}-\sqrt{15}} & \sqrt{3-\sqrt{5}} &
\sqrt{8+\sqrt{6 (3+\sqrt{5})}}
\end{array}
\right) \approx \left(
\begin{array}{ccc}
 0.896 & 0.387 & 0.219 \\
 0.309 & 0.896 & 0.319 \\
 0.319 & 0.219 & 0.922
\end{array}
\right)$&
$\begin{array}{c}(Z_{4},Z_{4})\\(Z_{4},Z_{12})\\(Z_{12},Z_{12})
\end{array}$  &
($f q h$, $a^2 f h q a$) & 0 &$\begin{array}{l}
\sin^2\theta_{12}\approx0.157\\ \sin^2\theta_{23}\approx0.107\\ \sin^2\theta_{13}\approx 0.048\,  \end{array}$ & 397.6\\
&&&&&&\\

IX &
$
\frac{1}{2 \sqrt{3 (3-\sqrt{5})}} \;
\left(
\begin{array}{ccc}
 \sqrt{11-\sqrt{3}-\sqrt{5} (3-\sqrt{3})} & \sqrt{11+\sqrt{3}-\sqrt{5} (3+\sqrt{3})} & 3-\sqrt{5}\\
  \sqrt{11+\sqrt{3}-\sqrt{5} (3+\sqrt{3})}  & \sqrt{11-\sqrt{3}-\sqrt{5} (3-\sqrt{3})}& 3-\sqrt{5}\\
3-\sqrt{5} & 3 -\sqrt{5} & 2 \sqrt{2}
\end{array}
\right)
\approx
\left(
\begin{array}{ccc}
 0.838 & 0.484 & 0.252 \\
 0.484 & 0.838 & 0.252 \\
 0.252 & 0.252 & 0.934
\end{array}
\right)$&  $\begin{array}{c}(Z_{4},Z_{3})\\(Z_{12},Z_{3})
\end{array}$  &
($f q h$, $q h a$)
& $\begin{array}{c}
\frac{1}{48} (\sqrt{5}-1)\\
\approx 0.0258
\end{array}$
& $\begin{array}{l}
\sin^2\theta_{12}\approx 0.251\, \\ \sin^2\theta_{23}\approx0.068\\ \sin^2\theta_{13}\approx 0.064\end{array}$ & 511.2\\
&&&&&&\\

Xa & $
\frac 16
\left(
\begin{array}{ccc}
 \sqrt{2 (9+\sqrt{5})} & 1+\sqrt{5} & 2 \sqrt{3-\sqrt{5}}\\
1+\sqrt{5} & 2 \sqrt{6-\sqrt{5}} & 1+\sqrt{5}\\
2 \sqrt{3-\sqrt{5}} & 1+\sqrt{5} &  \sqrt{2 (9+\sqrt{5})}
\end{array}
\right)
\approx
\left(
\begin{array}{ccc}
 0.790 & 0.539 & 0.291 \\
 0.539 & 0.647 & 0.539 \\
 0.291 & 0.539 & 0.790
\end{array}
\right)$& $\begin{array}{c}(Z_{3},Z_{3})
\end{array}$  &
($h f$, $h a q$)
& $\begin{array}{c}
\frac{1}{36 \sqrt{3}} (1+\sqrt{5})\\
\approx 0.0519
\end{array}$
& $\begin{array}{l}
\sin^2\theta_{12}\approx 0.318\, \star\\ \sin^2\theta_{23}\approx0.318\\ \sin^2\theta_{13}\approx 0.085\end{array}$ & 716.2\\
&&&&&&\\

Xb & $
\frac 16
\left(
\begin{array}{ccc}
 \sqrt{2 (9+\sqrt{5})} & 1+\sqrt{5} & 2 \sqrt{3-\sqrt{5}}\\
 2 \sqrt{3-\sqrt{5}} & 1+\sqrt{5} &  \sqrt{2 (9+\sqrt{5})}\\
1+\sqrt{5} & 2 \sqrt{6-\sqrt{5}} & 1+\sqrt{5}
\end{array}
\right)
\approx
\left(
\begin{array}{ccc}
 0.790 & 0.539 & 0.291 \\
 0.291 & 0.539 & 0.790\\
0.539 & 0.647 & 0.539
\end{array}
\right)$& $\begin{array}{c}(Z_{3},Z_{3})
\end{array}$  &
($h f$, $h a q$)
& $\begin{array}{c}
\frac{1}{36 \sqrt{3}} (1+\sqrt{5})\\
\approx 0.0519
\end{array}$
& $\begin{array}{l}
\sin^2\theta_{12}\approx 0.318\, \star\\ \sin^2\theta_{23}\approx0.682\\ \sin^2\theta_{13}\approx 0.085\end{array}$ &718.6 \\
&&&&&&\\

XIa & $
\frac 14 \;
\left(
\begin{array}{ccc}
 1+\sqrt{5} & 2 & -1+\sqrt{5}\\
2 & 2 \sqrt{2} & 2\\
-1+\sqrt{5} & 2 & 1+\sqrt{5}
\end{array}
\right)
\approx
\left(
\begin{array}{ccc}
 0.809 & 0.500 & 0.309 \\
 0.500 & 0.707 & 0.500 \\
 0.309 & 0.500 & 0.809
\end{array}
\right)$&  $\begin{array}{c}(Z_{4},Z_{4})\\(Z_{4},Z_{12})\\(Z_{12},Z_{12})\\ (Z_2\times Z_2, Z_2\times Z_2)
\end{array}$   &
$\begin{array}{c}
(f q h, q a f a^2)\\ \\ \\(\{ f q, h f q h\}, \{ a f a^2, a q a^2\})
\end{array}$
&  
 $\frac{\sqrt{3}}{32}
\approx 0.0541$
& $\begin{array}{l}
\sin^2\theta_{12}\approx 0.276\, \star\\ \sin^2\theta_{23}\approx0.276\\ \sin^2\theta_{13}\approx 0.095\end{array}$ & 993.5\\
&&&&&&\\

XIb & $
\frac 14 \;
\left(
\begin{array}{ccc}
 1+\sqrt{5} & 2 & -1+\sqrt{5}\\
 -1+\sqrt{5} & 2 & 1+\sqrt{5}\\
2 & 2 \sqrt{2} & 2
\end{array}
\right)
\approx
\left(
\begin{array}{ccc}
 0.809 & 0.500 & 0.309 \\
 0.309 & 0.500 & 0.809\\
0.500 & 0.707 & 0.500
\end{array}
\right)$&  $\begin{array}{c}(Z_{4},Z_{4})\\(Z_{4},Z_{12})\\(Z_{12},Z_{12})\\ (Z_2\times Z_2, Z_2\times Z_2)
\end{array}$   &
$\begin{array}{c}
(f q h, q a f a^2)\\ \\ \\(\{ f q, h f q h\}, \{ a f a^2, a q a^2\})
\end{array}$
&   
 $\frac{\sqrt{3}}{32}
\approx 0.0541$
& $\begin{array}{l}
\sin^2\theta_{12}\approx 0.276\,\star\\ \sin^2\theta_{23}\approx0.724\\ \sin^2\theta_{13}\approx 0.095\end{array}$ & 1000.0\\
&&&&&&\\

XII &
$ \frac{1}{2 \sqrt{15}}
\left(
\begin{array}{ccc}
\sqrt{25+\sqrt{5}+\sqrt{30 (5-\sqrt{5})}} & \sqrt{25+\sqrt{5}-\sqrt{30 (5-\sqrt{5})}} & \sqrt{2 (5-\sqrt{5})} \\
\sqrt{25+\sqrt{5}-\sqrt{30 (5-\sqrt{5})}} & \sqrt{25+\sqrt{5}+\sqrt{30 (5-\sqrt{5})}} & \sqrt{2 (5-\sqrt{5})} \\
 \sqrt{2 (5-\sqrt{5})} &  \sqrt{2 (5-\sqrt{5})} &  2 \sqrt{10+\sqrt{5}}
\end{array}
\right)
\approx
\left(
\begin{array}{ccc}
 0.778 & 0.550 & 0.304 \\
 0.550 & 0.778 & 0.304 \\
 0.304 & 0.304 & 0.903
\end{array}
\right)$&  $\begin{array}{c}(Z_{5},Z_{3})\\(Z_{15},Z_{3})
\end{array}$  &
($(h q a f q)^3$, $q h$)
& $\begin{array}{c}
\frac{1}{60} \sqrt{2 (5-\sqrt{5})}\\
\approx 0.0392
\end{array}$
 & $\begin{array}{l}
\sin^2\theta_{12}\approx 0.333 \, \star\\ \sin^2\theta_{23}\approx0.101\\ \sin^2\theta_{13}\approx 0.092\end{array}$ & 1034.3\\
&&&&&&\\

XIII & $
\frac{1}{\sqrt{12}} \;
\left(
\begin{array}{ccc}
\sqrt{5+\sqrt{15}} & \sqrt{2} & \sqrt{5-\sqrt{15}}\\
\sqrt{2} & 2 \sqrt{2}  & \sqrt{2} \\
 \sqrt{5-\sqrt{15}} & \sqrt{2} &  \sqrt{5+\sqrt{15}}
\end{array}
\right)
\approx
\left(
\begin{array}{ccc}
 0.860 & 0.408 & 0.306 \\
 0.408 & 0.816 & 0.408\\
 0.306 & 0.408 & 0.860
\end{array}
\right)$&  $\begin{array}{c}(Z_{4},Z_{3})\\(Z_{12},Z_{3})
\end{array}$    &
($f q h$, $a f h f a$)
&  
$\frac{1}{24}
\approx 0.0417$
 &  $\begin{array}{l}
\sin^2\theta_{12}\approx 0.184\,\\ \sin^2\theta_{23}\approx0.184\\ \sin^2\theta_{13}\approx 0.094\end{array}$& 1091.6\\
&&&&&&\\

XIV & $
\frac{1}{\sqrt{6 (3-\sqrt{5})}} \;
\left(
\begin{array}{ccc}
\sqrt{2} & \sqrt{2} & 3-\sqrt{5}\\
\sqrt{8-3\sqrt{5} +\sqrt{3 (9 -4\sqrt{5})}} & \sqrt{8-3\sqrt{5} -\sqrt{3 (9 -4\sqrt{5})}}  & \sqrt{2}\\
\sqrt{8-3\sqrt{5} -\sqrt{3 (9 -4\sqrt{5})}} & \sqrt{8-3\sqrt{5} +\sqrt{3 (9 -4\sqrt{5})}}  & \sqrt{2}
\end{array}
\right)
\approx
\left(
\begin{array}{ccc}
 0.661 & 0.661 & 0.357 \\
 0.609 & 0.439 & 0.661\\
 0.439 & 0.609 & 0.661
\end{array}
\right)$& $\begin{array}{c}(Z_{4},Z_{3})\\(Z_{12},Z_{3})
\end{array}$  &
($f q h$, $h q a$)
& $\begin{array}{c}
\frac{1}{48} (1+\sqrt{5})\\
\approx 0.0674
\end{array}$
 & $\begin{array}{l}
\sin^2\theta_{12}= 0.500\\ \sin^2\theta_{23}= 0.500\, \star\\ \sin^2\theta_{13}\approx 0.127\end{array}$ & 2238.5\\
&&&&&&\\

XV & $
\frac{1}{\sqrt{6 (5+\sqrt{5})}}
\left(
\begin{array}{ccc}
 \sqrt{2 (6 +\sqrt{5} + \sqrt{15 + 6 \sqrt{5}})} &  1+\sqrt{5} & \sqrt{2 (6 +\sqrt{5} - \sqrt{15 + 6 \sqrt{5}})} \\
1+\sqrt{5} & \sqrt{2 (9+\sqrt{5})} & 1+\sqrt{5}\\
\sqrt{2 (6 +\sqrt{5} - \sqrt{15 + 6 \sqrt{5}})}  &  1+\sqrt{5} &  \sqrt{2 (6 +\sqrt{5} + \sqrt{15 + 6 \sqrt{5}})}
\end{array}
\right)
\approx
\left(
\begin{array}{ccc}
 0.791 & 0.491 & 0.366 \\
 0.491 & 0.719 & 0.491 \\
 0.366 & 0.491 & 0.791
\end{array}
\right)$&  $\begin{array}{c}(Z_{5},Z_{3})\\(Z_{15},Z_{3})
\end{array}$  &
($(h q a f q)^3$, $h a q$)
& $\begin{array}{c}
\frac{1}{60} \sqrt{2 (5+\sqrt{5})}\\
 \approx 0.0634
 \end{array}$
 & $\begin{array}{l}
\sin^2\theta_{12}\approx 0.278 \, \star \\ \sin^2\theta_{23}\approx 0.278\\ \sin^2\theta_{13}\approx 0.134\end{array}$ & 2269.1\\
&&&&&&\\

XVI & $ \frac{1}{\sqrt{2 (5+\sqrt{5})}} \; \left(
\begin{array}{ccc}
 \sqrt{4 +\sqrt{5} + \sqrt{5+2 \sqrt{5}}} &  \sqrt{4 +\sqrt{5} - \sqrt{5+2 \sqrt{5}}} & \sqrt{2}\\
 \sqrt{4 +\sqrt{5} - \sqrt{5+2 \sqrt{5}}} &  \sqrt{4 +\sqrt{5} + \sqrt{5+2 \sqrt{5}}} & \sqrt{2}\\
 \sqrt{2} & \sqrt{2} & 1+\sqrt{5}
\end{array}
\right) \approx \left(
\begin{array}{ccc}
 0.802 & 0.467 & 0.372 \\
 0.467 & 0.802 & 0.372 \\
 0.372 & 0.372 & 0.851
\end{array}
\right)$&
$\begin{array}{c}(Z_{5},Z_{4})\\(Z_{15},Z_{4})\\(Z_{5},Z_{12})\\(Z_{15},Z_{12})
\end{array}$  &
($(h q a f q)^3$, $a^2 f q h a$) & $\begin{array}{c}
\frac{1}{80} \sqrt{6 (5-\sqrt{5})}\\
\approx 0.0509
\end{array}$
&  $\begin{array}{l}
\sin^2\theta_{12}\approx 0.253\,\\ \sin^2\theta_{23}\approx0.160\\ \sin^2\theta_{13}\approx 0.138\end{array}$& 2532.8\\
&&&&&&\\
 \hline 
\end{tabular}}
\end{center}
\end{table}
\end{landscape}

\begin{landscape}
\begin{table}
\caption{\label{tab:Sigma360x3Pattern3} Mixing patterns associated with  
$\Sigma(360\times3)$. Patterns called Ki require that one of the groups $G_e$ or $G_\nu$ is
a Klein group. Notice that only pattern KI requires $G_e$ to be a Klein group instead of $G_\nu$. The circle ($\circ$) next to the result of $|J_{CP}|$
for pattern XVIII indicates that this value corresponds to maximal CP violation $|\sin \delta|=1$.
For further explanation see table \ref{tab:Sigma360x3Pattern1}.}
\vspace{-20pt}
\begin{center}
\renewcommand{\tabcolsep}{.1mm}
 \renewcommand{\arraystretch}{.7}
{\scriptsize\tiny
\begin{tabular}{|c|lccccc|}
\hline

&&&&&&\\
\# & Pattern & ($G_{e}, G_{\nu}$)  & Example of ($g_{e},
g_{\nu}$)&   $|J_{CP}|$  & ($\sin^2\theta_{12}, \sin^2\theta_{23}, \sin^2\theta_{13}$) & $\quad\chi^2\quad$ \\
&&&&&&\\
\hline 

&&&&&&\\

XVII &
$\frac{1}{2 \sqrt{5+\sqrt{5}}} \;
\left(
\begin{array}{ccc}
\sqrt{7 + \sqrt{5} + \sqrt{2 (5+\sqrt{5})}} & 1+\sqrt{5} & \sqrt{7 + \sqrt{5} - \sqrt{2 (5+\sqrt{5})}} \\
1+\sqrt{5} & 2 \sqrt{2} & 1+\sqrt{5}\\
 \sqrt{7 + \sqrt{5} - \sqrt{2 (5+\sqrt{5})}} & 1+\sqrt{5} & \sqrt{7 + \sqrt{5} + \sqrt{2 (5+\sqrt{5})}}
\end{array}
\right)
\approx
\left(
\begin{array}{ccc}
 0.671 & 0.602& 0.433 \\
 0.602 & 0.526& 0.602 \\
 0.433 & 0.602 & 0.671
\end{array}
\right)$&  $\begin{array}{c}(Z_{5},Z_{4})\\(Z_{15},Z_{4})\\(Z_{5},Z_{12})\\(Z_{15},Z_{12})
\end{array}$   &
($(h q a f q)^3$, $a f h q a^2$)
&
$\begin{array}{c}
\frac{1}{80} \sqrt{6 (5+\sqrt{5})}\\
\approx 0.0824
\end{array}$
 & $\begin{array}{l}
\sin^2\theta_{12}\approx 0.445\\ \sin^2\theta_{23}\approx0.445\,\star\\ \sin^2\theta_{13}\approx 0.188\end{array}$ & 5060.1\\
&&&&&&\\

XVIII & $
\frac{1}{\sqrt{5}} \;
\left(
\begin{array}{ccc}
 \sqrt{3} & 1 & 1 \\
 1 & \sqrt{2} & \sqrt{2}\\
 1 & \sqrt{2} & \sqrt{2}
\end{array}
\right)
\approx
\left(
\begin{array}{ccc}
 0.775 & 0.447 & 0.447\\
 0.447 & 0.632& 0.632 \\
 0.447 & 0.632 & 0.632
\end{array}
\right)$&  $\begin{array}{c}(Z_{5},Z_{5})\\(Z_{5},Z_{15})\\(Z_{15},Z_{15})
\end{array}$  &
($(h q a f q)^3$, $q h a f q$)
&
$\frac{\sqrt{15}}{50}
\approx 0.0775 \; \circ$
& $\begin{array}{l}
\sin^2\theta_{12}= 0.250\\ \sin^2\theta_{23}=0.500\,\star\\ \sin^2\theta_{13}= 0.200\end{array}$ & 5715.1\\

&&&&&&\\
\hline

&&&&&&\\

KI & $
\frac 14 \; \left(
\begin{array}{ccc}
 1+\sqrt{5} & \sqrt{5+\sqrt{15}- \sqrt{2 (4+\sqrt{15})}} & \sqrt{5+\sqrt{3}-\sqrt{5} (1+\sqrt{3})}\\
 2 & \sqrt{2 (3+\sqrt{3})} & \sqrt{2 (3-\sqrt{3})} \\
 -1+\sqrt{5} &  \sqrt{5-\sqrt{3}+\sqrt{5} (1-\sqrt{3})} & \sqrt{5+\sqrt{15}+ \sqrt{2 (4+\sqrt{15})}}
\end{array}
\right)
\approx
\left(
\begin{array}{ccc}
 0.809 & 0.554 & 0.197 \\
 0.500 & 0.769 & 0.398 \\
 0.309 & 0.319 & 0.896
\end{array}
\right)$& $\begin{array}{c}(Z_{2}\times Z_{2},Z_4)\\(Z_{2}\times Z_{2},Z_{12})
\end{array}$
&  ($\{ f, q \}$, $a f q h a^2$)
 & $\frac{1}{32} \approx 0.0313$  & $\begin{array}{l}
\sin^2\theta_{12}\approx 0.319\,\star\\ \sin^2\theta_{23}\approx0.165\\ \sin^2\theta_{13}\approx 0.039\end{array}$ & 148.1\\
&&&&&&\\

KII & $
\frac{1}{\sqrt{2 (5+\sqrt{5})}} \; \left(
\begin{array}{ccc}
\sqrt{4+\sqrt{5} + \sqrt{15 + 6 \sqrt{5}}} & \sqrt{2} & \sqrt{4+\sqrt{5}-\sqrt{15+6 \sqrt{5}}}\\
\sqrt{2} & 1+\sqrt{5} & \sqrt{2}\\
\sqrt{4+\sqrt{5}-\sqrt{15+6 \sqrt{5}}} & \sqrt{2} & \sqrt{4+\sqrt{5} + \sqrt{15 + 6 \sqrt{5}}}
\end{array}
\right)
\approx
\left(
\begin{array}{ccc}
 0.894 & 0.372 & 0.250 \\
 0.372 & 0.851 & 0.372 \\
 0.250 & 0.372 & 0.894
\end{array}
\right) $& $\begin{array}{c}(Z_5, Z_{2}\times Z_{2})\\(Z_{15}, Z_{2}\times Z_{2})
\end{array}$  &
($(h q a f q)^3$, $\{ a^2 f a, a^2 q a \}$)
 &
 $\begin{array}{c}
 \frac{1}{80} \sqrt{2 (5 - \sqrt{5})}\\
\approx 0.0294
\end{array}$
 & $\begin{array}{l}
\sin^2\theta_{12}\approx 0.147\\ \sin^2\theta_{23}\approx0.147\,\\ \sin^2\theta_{13}\approx 0.063\end{array}$ & 553.5\\
&&&&&&\\

KIII & $
\frac{\sqrt{2}}{\sqrt{3} (1+\sqrt{5})} \; \left(
\begin{array}{ccc}
\sqrt{7 + 3 \sqrt{5}} & 1 & 1\\
1 & \sqrt{7 + 3 \sqrt{5}} & 1\\
1 & 1 & \sqrt{7 + 3 \sqrt{5}}
\end{array}
\right)
\approx
\left(
\begin{array}{ccc}
0.934 & 0.252 & 0.252 \\
0.252& 0.934 & 0.252\\
0.252 & 0.252 & 0.934
\end{array}
\right)$&
$(Z_3, Z_{2}\times Z_{2})$
&
($h a f a$, $\{ f, q \}$)
 &
 $\begin{array}{c}
 \frac{1}{144} \sqrt{6 (3- \sqrt{5})}\\
\approx 0.0149
\end{array}$
 & $\begin{array}{l}
\sin^2\theta_{12}\approx 0.068\\ \sin^2\theta_{23}\approx0.068\,\\ \sin^2\theta_{13}\approx 0.064\end{array}$ & 844.7\\
&&&&&&\\

KIVa & $
\frac{1}{2 \sqrt{5+\sqrt{5}}} \; \left(
\begin{array}{ccc}
  \sqrt{7+\sqrt{5}+\sqrt{6 (5+\sqrt{5})}} & 1+\sqrt{5} &   \sqrt{7+\sqrt{5}-\sqrt{6 (5+\sqrt{5})}} \\
  \sqrt{7+\sqrt{5}-\sqrt{6 (5+\sqrt{5})}}  & 1+\sqrt{5} &   \sqrt{7+\sqrt{5}+\sqrt{6 (5+\sqrt{5})}}\\
 1+\sqrt{5} & 2 \sqrt{2} & 1+\sqrt{5}
\end{array}
\right)
\approx
\left(
\begin{array}{ccc}
 0.739 & 0.602 & 0.302 \\
 0.302 & 0.602 & 0.739 \\
 0.602 & 0.526 & 0.602
\end{array}
\right) $& $\begin{array}{c}(Z_5, Z_{2}\times Z_{2})\\(Z_{15}, Z_{2}\times Z_{2})
\end{array}$  &
($(h q a f q)^3$, $\{ a f a^2, a q a^2 \}$)
 &
 $\begin{array}{c}
 \frac{1}{80} \sqrt{2 (5+ \sqrt{5})}\\
\approx 0.0476
\end{array}$
 & $\begin{array}{l}
\sin^2\theta_{12}\approx 0.398\\ \sin^2\theta_{23}\approx0.602\,\star\\ \sin^2\theta_{13}\approx 0.091\end{array}$ & 915.3\\
&&&&&&\\

KIVb & $
\frac{1}{2 \sqrt{5+\sqrt{5}}} \; \left(
\begin{array}{ccc}
  \sqrt{7+\sqrt{5}+\sqrt{6 (5+\sqrt{5})}} & 1+\sqrt{5} &   \sqrt{7+\sqrt{5}-\sqrt{6 (5+\sqrt{5})}} \\
   1+\sqrt{5} & 2 \sqrt{2} & 1+\sqrt{5}\\
  \sqrt{7+\sqrt{5}-\sqrt{6 (5+\sqrt{5})}}  & 1+\sqrt{5} &   \sqrt{7+\sqrt{5}+\sqrt{6 (5+\sqrt{5})}}
\end{array}
\right)
\approx
\left(
\begin{array}{ccc}
 0.739 & 0.602 & 0.302 \\
 0.602 & 0.526 & 0.602\\
0.302 & 0.602 & 0.739
\end{array}
\right) $&
 $\begin{array}{c}(Z_5, Z_{2}\times Z_{2})\\(Z_{15}, Z_{2}\times Z_{2})
\end{array}$
  &
  ( $(h q a f q)^3$, $\{ a f a^2, a q a^2 \}$)
   &
    $\begin{array}{c}
 \frac{1}{80} \sqrt{2 (5+\sqrt{5})}\\
\approx 0.0476
\end{array}$
   & $\begin{array}{l}
\sin^2\theta_{12}\approx 0.398\\ \sin^2\theta_{23}\approx0.398\,\star\\ \sin^2\theta_{13}\approx 0.091\end{array}$ & 915.5\\
&&&&&&\\

KV & $
\frac{1}{\sqrt{3 (3-\sqrt{5})}} \; \left(
\begin{array}{ccc}
 1 & 1 & \sqrt{7-3 \sqrt{5}}\\
 1 & \sqrt{7-3 \sqrt{5}} & 1\\
 \sqrt{7-3 \sqrt{5}} & 1 & 1
\end{array}
\right)
\approx
\left(
\begin{array}{ccc}
 0.661 & 0.661 & 0.357 \\
 0.661 & 0.357 & 0.661 \\
 0.357 & 0.661 & 0.661
\end{array}
\right) $&
$(Z_3, Z_{2}\times Z_{2})$
&
($a f h a^2$, $\{ f , q  \}$)
&
$\begin{array}{c}
\frac{1}{144} \sqrt{6 (3+\sqrt{5})}\\
\approx0.0389
\end{array}$
& $\begin{array}{l}
\sin^2\theta_{12}= 0.500\\ \sin^2\theta_{23}=0.500\,\star\\ \sin^2\theta_{13}\approx 0.127\end{array}$ & 2238.5\\
&&&&&&\\

KVI & $
\frac{1}{2 \sqrt{2}} \; \left(
\begin{array}{ccc}
\sqrt{3+\sqrt{3}} & \sqrt{2} & \sqrt{3-\sqrt{3}}\\
\sqrt{2} & 2 & \sqrt{2}\\
\sqrt{3-\sqrt{3}} & \sqrt{2} & \sqrt{3+\sqrt{3}}
\end{array}
\right)
\approx
\left(
\begin{array}{ccc}
 0.769 & 0.500 & 0.398 \\
 0.500 & 0.707 & 0.500 \\
 0.398 & 0.500 & 0.769
\end{array}
\right)$& $\begin{array}{c}(Z_4, Z_{2}\times Z_{2})\\(Z_{12}, Z_{2}\times Z_{2})
\end{array}$  &
($f q h$, $\{ f , a f a^2  \}$)
& $\frac{\sqrt{5}}{32} \approx0.0699$ & $\begin{array}{l}
\sin^2\theta_{12}\approx 0.297\,\star\\ \sin^2\theta_{23}\approx0.297\\ \sin^2\theta_{13}\approx 0.158\end{array}$ & 3360.7\\
&&&&&&\\

KVII & $
\frac{1}{\sqrt{6}} \; \left(
\begin{array}{ccc}
 2 & 1 & 1\\
 1 & 2 & 1\\
 1 & 1 & 2
\end{array}
\right)
\approx
\left(
\begin{array}{ccc}
 0.816 & 0.408& 0.408 \\
 0.408 & 0.816 & 0.408 \\
 0.408 & 0.408 & 0.816
\end{array}
\right)$&
$(Z_3, Z_{2}\times Z_{2})$
 &
 ( $a$, $\{ f q, h f q h \}$)
 & $\frac{\sqrt{15}}{72} \approx0.0538$ & $\begin{array}{l}
\sin^2\theta_{12}= 0.200\,\\ \sin^2\theta_{23}=0.200\\ \sin^2\theta_{13}\approx 0.167\end{array}$ & 3892.3\\

&&&&&&\\ \hline 
\end{tabular}}
\end{center}
\end{table}
\end{landscape}


\normalsize
As regards the vector which is of the same form (in absolute values) as the second column of the pattern KIVa we notice that it has been discussed as mixing vector in \cite{groupscan1}. There it has been
 associated with the group $\Sigma (60)$ which is isomorphic to $A_5$. Due to the relation between the representation matrices of $Z_2$ and $Z_6$ generating elements (via the center of $SU(3)$) the sets of mixing vectors have to coincide.

\section{Summary}
\label{concl}

Non-abelian discrete flavor symmetries $G_l$ which are broken to subgroups $G_e$ and $G_\nu$ in the charged lepton and neutrino sector, respectively, are still considered to be an interesting 
possibility to explain the peculiar lepton mixing pattern. 
The discovery of a large reactor mixing angle $\theta_{13} \approx 0.15$, however, strongly disfavors many mixing patterns such as TB mixing,
which is related to the flavor group $S_4$.
 Thus,  efforts have been made to find new symmetries, usually larger than the group $S_4$, which could predict the experimentally observed mixing parameters.
We have pursued such an attempt in the present paper and have discussed
 the ``exceptional" finite groups $\Sigma (n \varphi)$, $\varphi=3$, which are subgroups of $SU(3)$. These four groups are interesting, since  
 they all contain irreducible faithful three-dimensional representations and have several types of abelian subgroups, suitable as $G_e$ and $G_\nu$. 
 Since one of them, the group $\Sigma (360 \times 3)$, also comprises Klein subgroups, neutrinos could be Majorana particles as well.

Having in mind a realization of the mixing patterns in a scenario with only small corrections to the leading order results, we find that only a few of our patterns can give a good fit to the experimental data: 
one pattern associated with the group $\Sigma (36 \times 3)$, see eq.(\ref{S36x3_2}), which
leads to $\theta_{13} \approx 0.18$, one associated with $\Sigma (216 \times 3)$ with $\theta_{13} \approx 0.12$, see table \ref{tab:Sigma216x3Pattern1},\footnote{In our analysis of $\Sigma (216 \times 3)$ the pattern 
which fits the data best, see eq.(\ref{S216x3_add}), is not associated with $\Sigma (216 \times 3)$, but only with one of its subgroups of order 162. Also this pattern predicts a trivial Dirac phase.} 
as well as two patterns associated with $\Sigma (360 \times 3)$ which give rise to $\theta_{13} \approx 0.20$ and $\theta_{13} \approx 0.19$, respectively, see table \ref{tab:Sigma360x3Pattern1}.
 Although they are associated with $\Sigma (360 \times 3)$, also the latter two patterns require neutrinos to be Dirac particles.
  Interestingly enough, all these patterns lead to vanishing $J_{CP}$. A non-trivial CP phase only arises from
 patterns that do not accommodate the data well, i.e. the minimum value of $\chi^2$ is larger than 100, see table \ref{tab:Sigma216x3Pattern1}, \ref{tab:Sigma216x3Pattern2} 
 and \ref{tab:Sigma360x3Pattern1}-\ref{tab:Sigma360x3Pattern3}. Notice that among these are all patterns which can be derived for Majorana neutrinos from $G_l=\Sigma (360 \times 3)$.
  If they are implemented in a concrete model, large corrections to the pattern, derived from the breaking of $G_l$ to $G_e$ and to $G_\nu$, have to be present so that
  the mixing angles agree reasonably well with the data. Another possibility could be to modify the breaking pattern, for example by reducing one of the residual symmetries $G_e$ or $G_\nu$
  or by involving CP symmetry in the breaking. In this way, all the presented patterns could function as starting point for the search for new patterns which lead to mixing parameters agreeing well
  with the data.
   For the group $\Sigma (216 \times 3)$ we find several patterns with $|J_{CP}| \neq 0$ which allow 
 the solar and atmospheric mixing angles to be within the experimental $3 \, \sigma$ ranges, but lead to a too large value of $\theta_{13}$. It would be interesting to
 analyze whether a class of models can be constructed in which $\theta_{13}$ can be corrected appropriately, while the other two mixing angles (and the prediction of the Jarlskog invariant) only undergo small
 corrections. 
 
 Eventually, we have studied the outer automorphisms of the groups $\Sigma (n \varphi)$, $\varphi=3$, since we can relate inequivalent irreducible representations with their help.
In particular, we could show that all irreducible faithful three-dimensional representations of a group $\Sigma (n \varphi)$ lead to the same set of mixing patterns, and thus that it is sufficient to only 
consider one of these representations in our discussion.

\section*{Acknowledgements}

We thank Ferruccio Feruglio for encouragement. CH would like to thank Martin Holthausen and Michael A. Schmidt
for help with the computer program GAP as well as Roman Zwicky for email correspondence.
CH is supported by the ERC Advanced Grant no. 267985, ``Electroweak Symmetry Breaking, Flavour and Dark Matter: One Solution for Three Mysteries" (DaMeSyFla). CH thanks the Galileo Galilei Institute for Theoretical
Physics in Arcetri, Florence, for the hospitality and the INFN for partial support during the final stages of this work.
AM acknowledges partial support by the INFN program on ``Astroparticle Physics" and the Italian MIUR program on ``Neutrinos, Dark Matter and Dark Energy in the Era of LHC"
and thanks the University of Padua for hospitality and support during her visits.
The work of LV is supported by the Swiss National Science Foundation under the contract 200021-125237.

\appendix

\section{Comments on outer automorphisms of \mathversion{bold}$\Sigma (n \varphi)$\mathversion{normal}}
\label{app}

We would like to consider outer automorphisms of the groups $\Sigma (n \varphi)$, since these allow to relate inequivalent irreducible representations of a group with each other.
The group of outer automorphisms is defined as the group of all automorphisms divided by the inner ones. The latter are mappings represented by conjugation with an element
of the group. The connection between (outer) automorphisms of a group and symmetries of its character table is also discussed in \cite{flavorCP_2}.

\subsection{Outer automorphisms of \mathversion{bold}$\Sigma (36 \times 3)$ \mathversion{normal}}
\label{appS36x3}

The outer automorphisms of $\Sigma (36 \times 3)$ form a Klein group and thus three non-trivial mappings exist which act in general non-trivially on the inequivalent irreducible representations of the group. The first one can be defined on the 
generators of the group as
\be
a \;\; \rightarrow \;\; z^2 a \;\; , \;\;\; z \;\; \rightarrow \;\; z a \;\; , \;\;\; v \;\; \rightarrow \;\; a z v \; .
\ee  
It exchanges the representations ${\bf 3^{(0)}}$ and ${\bf 3^{(2)}}$ with their complex conjugates and ${\bf 3^{(1)}}$ with ${\bf (3^{(3)})^\star}$ as well as ${\bf 3^{(3)}}$ with ${\bf (3^{(1)})^\star}$.
This transformation is a symmetry of the character table, if we perform at the same time the following exchanges among the classes: $C_2 \; \leftrightarrow \; C_3$,
$C_7 \; \leftrightarrow \; C_8$, $C_{10} \; \leftrightarrow \; C_{11}$ and $C_{13} \; \leftrightarrow \; C_{14}$ (we refer here to the notation of the classes as found in \cite{Sigma}). The second
mapping acts on the generators as follows 
 \be
a \;\; \rightarrow \;\; z a^2 z \;\; , \;\;\; z \;\; \rightarrow \;\; z a^2 \;\; , \;\;\; v \;\; \rightarrow \;\; z^2 a^2 z v^3 \; .
\ee 
In terms of representations it interchanges the one-dimensional representations ${\bf 1^{(1)}}$ and ${\bf 1^{(3)}}$ as well as ${\bf 3 ^{(1)}}$ with ${\bf 3 ^{(3)}}$ (and consequently also ${\bf (3 ^{(1)})^\star}$ with ${\bf (3 ^{(3)})^\star}$).
It is a symmetry of the character table, if we exchange $C_9 \; \leftrightarrow \; C_{12}$, $C_{10} \; \leftrightarrow \; C_{13}$ and $C_{11} \; \leftrightarrow \; C_{14}$. The third mapping is found if the two presented ones
are performed one after another (the ordering is irrelevant, because they form a Klein group). It corresponds to an interchange of all representations with their complex conjugates.

\subsection{Outer automorphisms of \mathversion{bold}$\Sigma (72 \times 3)$ \mathversion{normal}}
\label{appS72x3}

The group of outer automorphisms of $\Sigma (72 \times 3)$
 is the permutation group $S_3$. Thus, we can define two mappings as generators. The first one acts as
\be
\label{eq:S72x3map1}
a \;\; \rightarrow \;\; z^2  \;\; , \;\;\; z \;\; \rightarrow \;\;  a^2 \;\; , \;\;\; v \;\; \rightarrow \;\; v^3 \;\; , \;\;\; x \;\; \rightarrow \;\; z^2 v^3 x
\ee
and thus exchanges the following representations: ${\bf 1^{(1,0)}}$ and ${\bf 1^{(1,1)}}$, ${\bf 3^{(1,0)}}$ with ${\bf (3^{(1,1)})^\star}$ and their complex conjugates as well as ${\bf 3^{(0,0)}}$, ${\bf 3^{(0,1)}}$ and ${\bf 6}$
with their complex conjugates. If we exchange the classes $C_2$ and $C_3$, $C_6$ and $C_7$, $C_9$ and $C_{10}$, $C_{11}$ and $C_{16}$, $C_{12}$ with $C_{15}$ and $C_{13}$ with $C_{14}$ (referring 
to the character table given in \cite{Sigma}), the character table remains invariant. Note this mapping generates a $Z_2$ symmetry. The second outer automorphism we can define sends the generators
$a$, $z$, $v$ and $x$ into
\be
\label{eq:S72x3map2}
a \;\; \rightarrow \;\; z a  \;\; , \;\;\; z \;\; \rightarrow \;\; a z a^2 \;\; , \;\;\; v \;\; \rightarrow \;\; z^2 a z v^3 x \;\; , \;\;\; x \;\; \rightarrow \;\; z a z a^2 v \; .
\ee
Its effect on the representations is: ${\bf 1^{(1,0)}} \;\rightarrow \; {\bf 1^{(1,1)}} \; \rightarrow \; {\bf 1^{(0,1)}} \;\rightarrow \; {\bf 1^{(1,0)}}$ and similarly
  ${\bf 3^{(1,0)}} \; \rightarrow \; {\bf 3^{(1,1)}} \; \rightarrow \; {\bf 3^{(0,1)}} \;\rightarrow \; {\bf 3^{(1,0)}}$ and the same holds for their complex conjugated representations. If we 
  cyclicly permute the classes $C_8 \; \rightarrow \; C_{16} \; \rightarrow \; C_{11} \; \rightarrow \; C_8$, $C_9 \; \rightarrow \; C_{14} \; \rightarrow \; C_{12} \; \rightarrow \; C_9$ 
  and $C_{10} \; \rightarrow \; C_{15} \; \rightarrow \; C_{13} \; \rightarrow \; C_{10}$ together with the representations,
  this mapping is a symmetry of the character table. Notice that it generates a $Z_3$ symmetry and that it does not commute with the action of the mapping defined in  eq.(\ref{eq:S72x3map1}). The latter is expected
  since the two mappings should give rise to the group $S_3$. Thus, there are in addition three non-trivial mappings: the map in eq.(\ref{eq:S72x3map2}) applied twice,
  the application of the first map followed by the second one as well as vice versa.  

\subsection{Outer automorphisms of \mathversion{bold}$\Sigma (216 \times 3)$ \mathversion{normal}}
\label{appS216x3}

The outer automorphism group is a $Z_6$ symmetry.  We can define a mapping of the generators as\footnote{Note that if applied only to the 
generators $a$, $v$ and $z$ which give rise to the group $\Sigma (36 \times 3)$ this mapping only generates a $Z_2$ symmetry instead of a $Z_6$.}
\be
a \;\; \rightarrow \;\; a^2 \;\; , \;\;\; z \;\; \rightarrow \;\; z \;\; , \;\;\; v \;\; \rightarrow \;\; v^3  \;\; , \;\;\; w \;\; \rightarrow \;\;  (z w)^2 \; .
\ee  
Applying this mapping we see that the three-dimensional representations and their complex conjugates transform as
\be
{\bf 3^{(0)}} \; \rightarrow \; {\bf (3^{(1)})^\star} \; , \;\; {\bf 3^{(1)}} \; \rightarrow \; {\bf (3^{(2)})^\star} \; , \;\; {\bf 3^{(2)}} \; \rightarrow \; {\bf (3^{(0)})^\star} \; .
\ee
The other representations are mapped as follows: ${\bf 1^{(1)}} \; \leftrightarrow \; {\bf 1^{(2)}}$, ${\bf 2^{(1)}} \; \leftrightarrow \; {\bf 2^{(2)}}$, ${\bf 6^{(0)}} \; \rightarrow \; {\bf (6^{(1)})^\star}$, ${\bf 6^{(1)}} \; \rightarrow \; {\bf (6^{(2)})^\star}$, 
${\bf 6^{(2)}} \; \rightarrow \; {\bf (6^{(0)})^\star}$  (and similarly their complex conjugates), as well as ${\bf 8^{(1)}} \; \leftrightarrow \; {\bf 8^{(2)}}$ and ${\bf 9} \; \leftrightarrow \; {\bf 9^\star}$.
Using the convention of \cite{Sigma} for the character table, we see that we have to exchange the classes $C_2 \; \leftrightarrow \; C_3$, $C_6 \; \leftrightarrow \; C_7$, $C_9 \; \leftrightarrow \; C_{10}$, 
$C_{11} \; \leftrightarrow \; C_{12}$ as well as apply the transformations $C_{13} \; \rightarrow \; C_{16} \; \rightarrow \; C_{15} \; \rightarrow \; C_{17} \; \rightarrow \; C_{14} \; \rightarrow \; C_{18} \; \rightarrow \; C_{13}$ and
$C_{19} \; \rightarrow \; C_{24} \; \rightarrow \; C_{21} \; \rightarrow \; C_{22} \; \rightarrow \; C_{20} \; \rightarrow \; C_{23} \; \rightarrow \; C_{19}$, if we want to promote this mapping to a symmetry of the character table. 
As one can check, applying this mapping repeatedly we can relate all faithful irreducible three-dimensional representations to ${\bf 3^{(0)}}$.

\normalsize


\end{document}